
\input psfig
\input mn.tex
\overfullrule=0pt


\def\vcirc{v_{\rm circ}}
\def\kms{\;{\rm kms}^{-1}}

\def\pc{\;\rm pc}
\def\min{\;\rm min}
\def\ds{\displaystyle}
\def\frac#1#2{\textstyle {#1\over #2} \displaystyle}

\def\curlyH{{\cal H}}
\def\curlyG{{\cal G}}
\def\lb{\frac{3}{2}\!+\!\frac{\gamma\!-\!5}{\alpha}}
\def\mlb{\frac{3}{2}\!+\!\frac{5\!-\!\gamma}{\alpha}}
\def\cmb{\frac{5\!-\!\gamma}{\alpha}\!-\!\frac{1}{2}}
\def\sigmap{\sigma_{\rm P}}
\def\los{{\rm los}}
%
%
\def\spose#1{\hbox to 0pt{#1\hss}}
\def\lta{\mathrel{\spose{\lower 3pt\hbox{$\sim$}}
    \raise 2.0pt\hbox{$<$}}}
\def\gta{\mathrel{\spose{\lower 3pt\hbox{$\sim$}}
    \raise 2.0pt\hbox{$>$}}}
\def\today{\ifcase\month\or
 January\or February\or March\or April\or May\or June\or
 July\or August\or September\or October\or November\or December\fi
 \space\number\day, \number\year}
\newdimen\hssize
\hssize=8.4truecm  
\newdimen\hdsize
\hdsize=17.7truecm    


\newcount\eqnumber
\eqnumber=1
\def\chaphead{}
 
\def\new{\hbox{(\rm\chaphead\the\eqnumber)}\global\advance\eqnumber by 1}
 
\def\first{\hbox{(\rm\chaphead\the\eqnumber a)}\global\advance\eqnumber by 1}
\def\last#1{\advance\eqnumber by -1 \hbox{(\rm\chaphead\the\eqnumber#1)}\advance
     \eqnumber by 1}
 
\def\ref#1{\advance\eqnumber by -#1 \chaphead\the\eqnumber
     \advance\eqnumber by #1}
 
\def\nref#1{\advance\eqnumber by -#1 \chaphead\the\eqnumber
     \advance\eqnumber by #1}

\def\eqnam#1{\xdef#1{\chaphead\the\eqnumber}}
 


\newcount\tabnumber 
\tabnumber=1
\def\tabnew{\global\advance\tabnumber by 1}
\def\tabnam#1{\xdef#1{\chaphead\the\tabnumber}}


\newcount\fignumber 
\fignumber=1
\def\fignew{\global\advance\fignumber by 1}
\def\fignam#1{\xdef#1{\chaphead\the\fignumber}}


\pageoffset{-0.85truecm}{-1.05truecm}



\pagerange{}
\pubyear{version: \today}
\volume{}


\begintopmatter

\title{Dark Matter in Dwarf Spheroidals I: Models}

\author{M.I.\ Wilkinson$^{1}$, J.\ Kleyna$^{1}$, N.W.\ Evans$^{2}$, G.\ Gilmore$^1$}

\vskip0.15truecm

\affiliation{$^1$ Institute of Astronomy, Madingley Rd, Cambridge CB3 0HA}

\vskip0.15truecm
\affiliation{$^2$ Theoretical Physics, Department of Physics, 1 Keble Road,
                 Oxford, OX1 3NP}

\shortauthor{M.I. Wilkinson, J. Kleyna, N.W. Evans \& G. Gilmore}

\shorttitle{Dwarf Spheroidals I: Models}



\abstract{This paper introduces a new two-parameter family of dwarf 
spheroidal (dSph) galaxy models. The mass distribution has a Plummer
profile and falls like $R^{-4}$ in projection in agreement with the
star-count data.  The first free parameter controls the velocity
anisotropy, the second controls the dark matter content. The dark
matter distribution can be varied from one extreme of
mass-follows-light through a near-isothermal halo with flat rotation
curve to the other extreme of an extended dark halo with harmonic
core.  This family of models is explored analytically in some detail
-- the distribution functions, the intrinsic moments and the projected
moments are all calculated.

For the nearby Galactic dSphs, samples of hundreds of discrete radial
velocities are becoming available. A technique is developed to extract
the anisotropy and dark matter content from such data sets by
maximising the likelihood function of the sample of radial
velocities. This is constructed from the distribution function and
corrected for observational errors and the effects of binaries. Tests
on simulated data sets show that samples of $\sim 1000$ discrete radial
velocities are ample to break the degeneracy between mass and
anisotropy in the nearby dSphs. Interesting constraints can already be
placed on the distribution of the dark matter with samples of $\sim
160$ radial velocities (the size of the present-day data set for
Draco).

The {\it Space Interferometry Mission} or SIM allows very accurate
differential astrometry at faint magnitudes. This can be used to
measure the internal proper motions of stars in the nearby Galactic
dSphs. Our simulations show that $\sim 100$ proper motions are
sufficient to demolish completely the mass-anisotropy degeneracy. The
target stars in Draco are at magnitudes of $V \sim 19-20$ and the
required proper motion accuracy is $3-6\,\mu{\rm as\,yr}^{-1}$. The
measurement of the proper motions of a sample of $\sim 100$ stars
uncontaminated with binaries will take about 400 hours of SIM time, or
under $2 \%$ of the mission lifetime.}


\maketitle  

%
\section{Introduction}
The dark matter content of low-luminosity dwarf galaxies, as inferred
from analyses of their internal stellar and gas kinematics, makes them
the most dark matter dominated of all galaxies (Mateo 1998, Carignan
\& Beaulieu 1989). Of these small galaxies, the low-luminosity
gas-free dwarf spheroidals (dSph) are the most extreme.  Available
stellar kinematic studies provide strong evidence for the presence of
dominant dark matter (e.g., Aaronson 1983, Mateo 1998), confirming
speculations based on estimates of the dSph's tidal radii (Faber \&
Lin 1983). Although there have been alternative suggestions as to the
origin of the large mass-to-light ratios of the dSphs (e.g., Kuhn \&
Miller 1989, Kroupa 1997), none of these have carried much conviction
-- see, for example, the objections raised by Sellwood \& Pryor (1997)
and Mateo (1997). The spatial distribution of the dark matter in the
dSphs is very poorly known. Three obvious possibilities suggest
themselves. First, the dark matter may shadow the stars and so the
mass distribution may follow the light. Second, the dark matter may be
distributed in a halo which generates a flat rotation curve, as is the
case for galaxies like the Milky Way or M31. Third, the scale length
of the dark matter may be larger than the luminous matter and so the
dSph may lie in the harmonic core of an extended dark matter halo. One
promising way to distinguish between these is by dynamical modelling.

Amongst the nearest dSphs are Draco and Sculptor, at heliocentric
distances of $82$ and $79$ kpc, respectively. Both are very attractive
candidates for the study of dark matter through dynamical modelling
being relatively simple systems with evidence for substantial dark
matter content. The Draco dSph has an inferred central mass-to-light
ratio of $\sim 60$ in solar V-band units, while Sculptor has a less
extreme value of $\sim 10$ (Mateo 1998).  Any robust dynamical
analysis is eased if the potential is approximately steady-state, and
if the tracer stellar distribution is in equilibrium and
well-mixed. The internal crossing times of the dSphs are typically
only
$$ t_{\rm cross} \sim
R/\sigma \sim 2 \times 10^7 (R/200{\pc})(10\kms /\sigma)\, {\rm yr}.
\eqno\new$$
The Draco dSph is dominated by an intermediate-age to old stellar
population, but with little star formation for the last $\sim 5$~Gyr
(Hernandez, Gilmore \& Valls-Gabaud 2000).  Deep HST imaging of a
small field in the outer regions of the Sculptor dSph has revealed
a stellar population that is old and metal-poor, similar to halo
globular clusters, with little spread in age (Monkiewicz et al. 1999)
or metallicity (Mateo 1998). Thus, star formation indeed ceased many
crossing times ago and both these systems should be well-mixed.
Radial velocity surveys are available for both Draco (Hargreaves et
al. 1996; Armandroff et al. 1995) and Sculptor (Queloz, Dubath \&
Pasquini 1995). Sculptor is of particular current interest due to the
detection of significant amounts of HI gas projected within its tidal
radius, at a consistent velocity to be truly associated with the dSph
(Carignan et al.\ 1998).

The aim of this paper is to provide new models for the dSphs and new
techniques for probing the dark matter distribution. Section 2
describes our models and their intrinsic properties, while Section 3
presents the observable properties, including the distributions of
radial velocities. In Section 4, Monte Carlo simulations are used to
assess how radial velocity surveys and proper motions inferred from
astrometric satellites discriminate between different dark matter
distributions. There are already samples of over a hundred radial
velocities available for Draco, and this will rise to several hundreds
in the next few years. The {\it Space Interferometry Mission} (SIM,
see ``http://sim.jpl.nasa.gov/'') has the capabilities to measure the
internal proper motions of stars in Draco. We assess the likely impact
of the new data sets and devise strategies for exploiting them. Finally,
a companion paper in this issue of {\it Monthly Notices} presents an
application of the models and algorithm to a newly acquired data set
for Draco (Kleyna et al. 2001, henceforth Paper II)

\begintable*{\tabnumber}
\tabnam{\constants}
\caption{{\bf Table \constants.} The numerical constants in the DFs. 
These are pure numbers fixed once and for all by the choice of the
anisotropy parameter $\gamma$ and the dark matter parameter $\alpha$.}
\halign{#\hfil&\quad#\hfil&\quad#\hfil\cr
\noalign{\hrule}
\noalign{\vskip0.3truecm}
$\alpha > 0$ & $L^2 \le 2E$ & $C_{\alpha,\gamma} =
{\ds\rho_0\over \ds\psi_0^{5/\alpha -\gamma/\alpha}}
           {\ds \Gamma (5/\alpha - \gamma/\alpha +1) \over
           \ds (2\pi)^{3/2} \Gamma (5/\alpha - \gamma/\alpha -
1/2)}$\cr
$\alpha > 0$ & $L^2 \ge 2E$ & $C_{\alpha,\gamma} =
{\ds\rho_0\over \ds\psi_0^{5/\alpha -\gamma/\alpha}}
           {\ds \Gamma (5/\alpha - \gamma/\alpha +1) \over
           \ds (2\pi)^{3/2} \Gamma ( 1 - \gamma/2)
          \Gamma (5/\alpha + \gamma/2 - \gamma/\alpha - 1/2)}$\cr
\null & \null & \null \cr
$\alpha = 0$ & $\forall L^2, E$ & $C_{0,\gamma} = {\ds \rho_0 \over \ds v_0^3} \Bigl(
{\ds (5-\gamma) \over \ds 2 \pi} \Bigr)^{3/2} $\cr
\null & \null & \null \cr
$\alpha < 0$ & $L^2 \le 2E$ & $C_{\alpha,\gamma} =
{\ds\rho_0\over \ds (-\psi_0)^{5/\alpha -\gamma/\alpha}}
           {\ds \Gamma (\gamma/\alpha - 5/\alpha + 3/2) \over
           \ds (2\pi)^{3/2} \Gamma (\gamma/\alpha - 5/\alpha)}$\cr
$\alpha < 0$ & $L^2 \ge 2E$ & $C_{\alpha,\gamma} =
{\ds\rho_0\over \ds (-\psi_0)^{5/\alpha -\gamma/\alpha}}
           {\ds \Gamma (\gamma/\alpha - 5/\alpha -\gamma /2 + 3/2) \over
           \ds (2\pi)^{3/2} \Gamma(1-\gamma/2)
           \Gamma (\gamma/\alpha - 5/\alpha)}$\cr
\noalign{\vskip0.3truecm}
\noalign{\hrule}
}
\tabnew
\endtable
\beginfigure*{\fignumber}
\fignam{\figstarcount}
\centerline{\psfig{figure=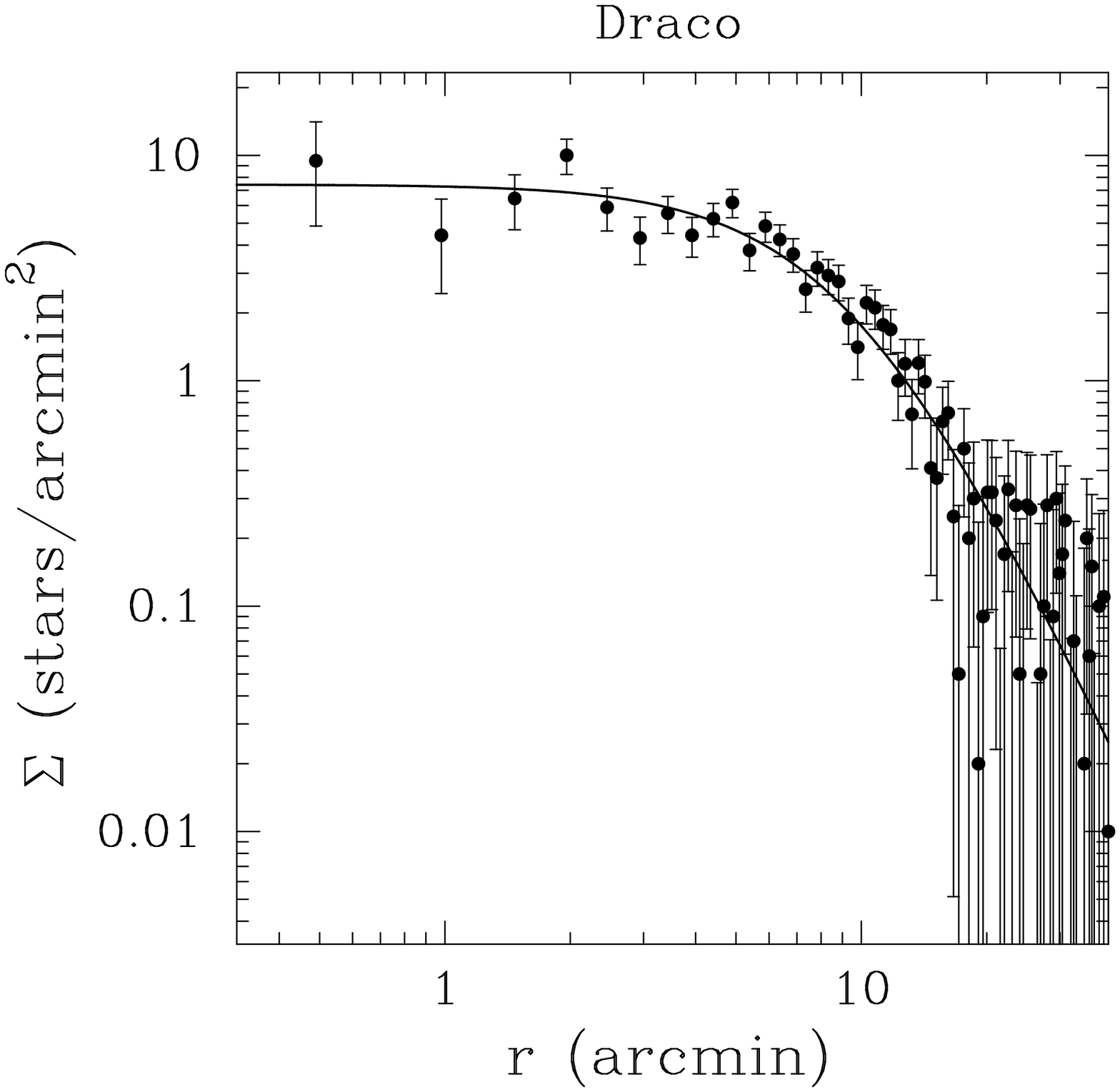,width=\hssize}}
\centerline{\psfig{figure=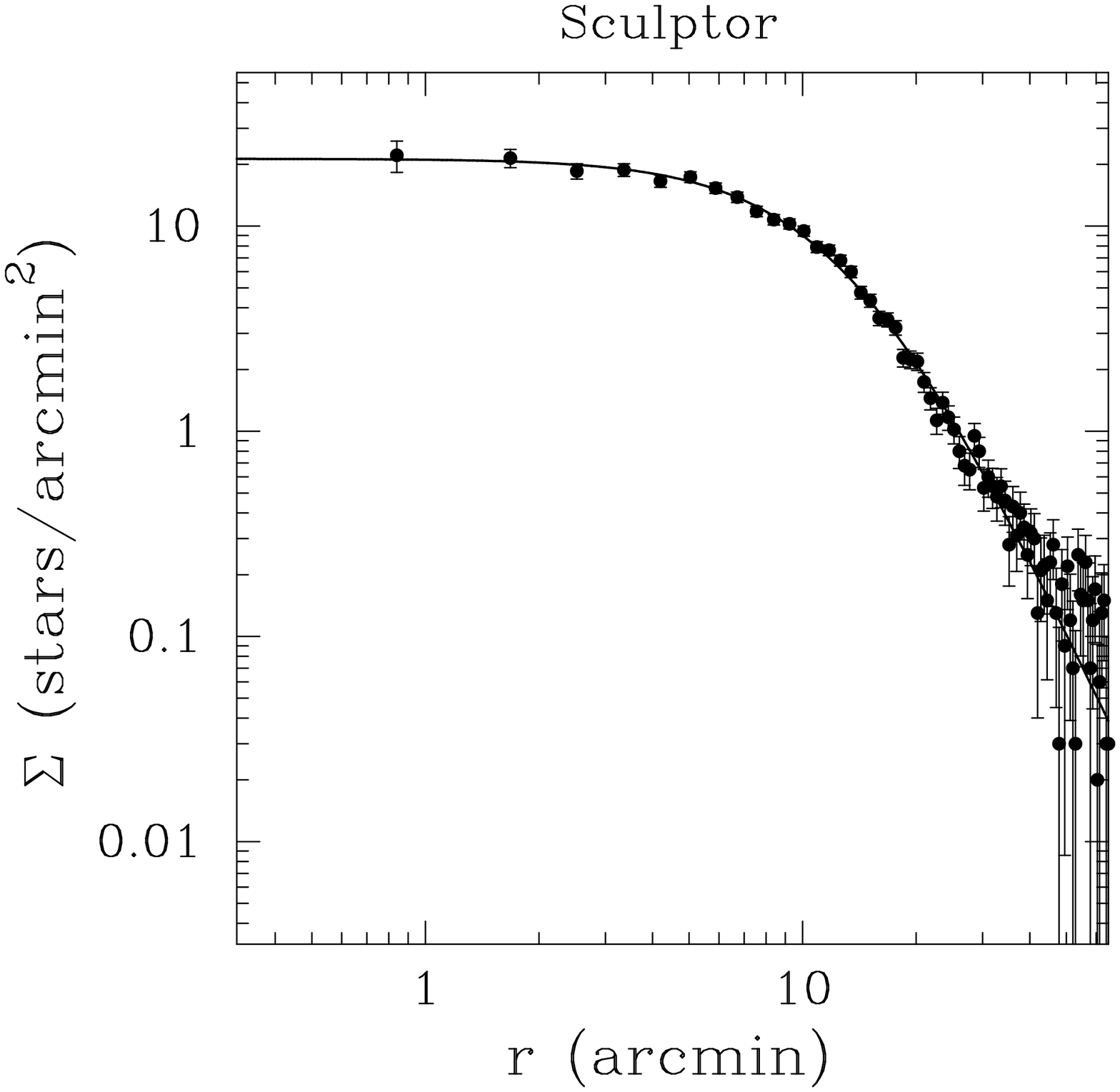,width=\hssize}}
\smallskip\noindent
\caption{{\bf Figure \figstarcount.} This shows the projected density
of the best fitting Plummer model for Draco (upper panel) and Sculptor
(lower panel). The data are the background-subtracted star count major
axis profiles given by Irwin \& Hatzidimitriou (1995). ($r_0 =9.71$
arcmin for Draco and $13.64$ arcmin for Sculptor).}
\fignew
\endfigure
\beginfigure*{\fignumber}
\fignam{\masstolight}
\centerline{\psfig{figure=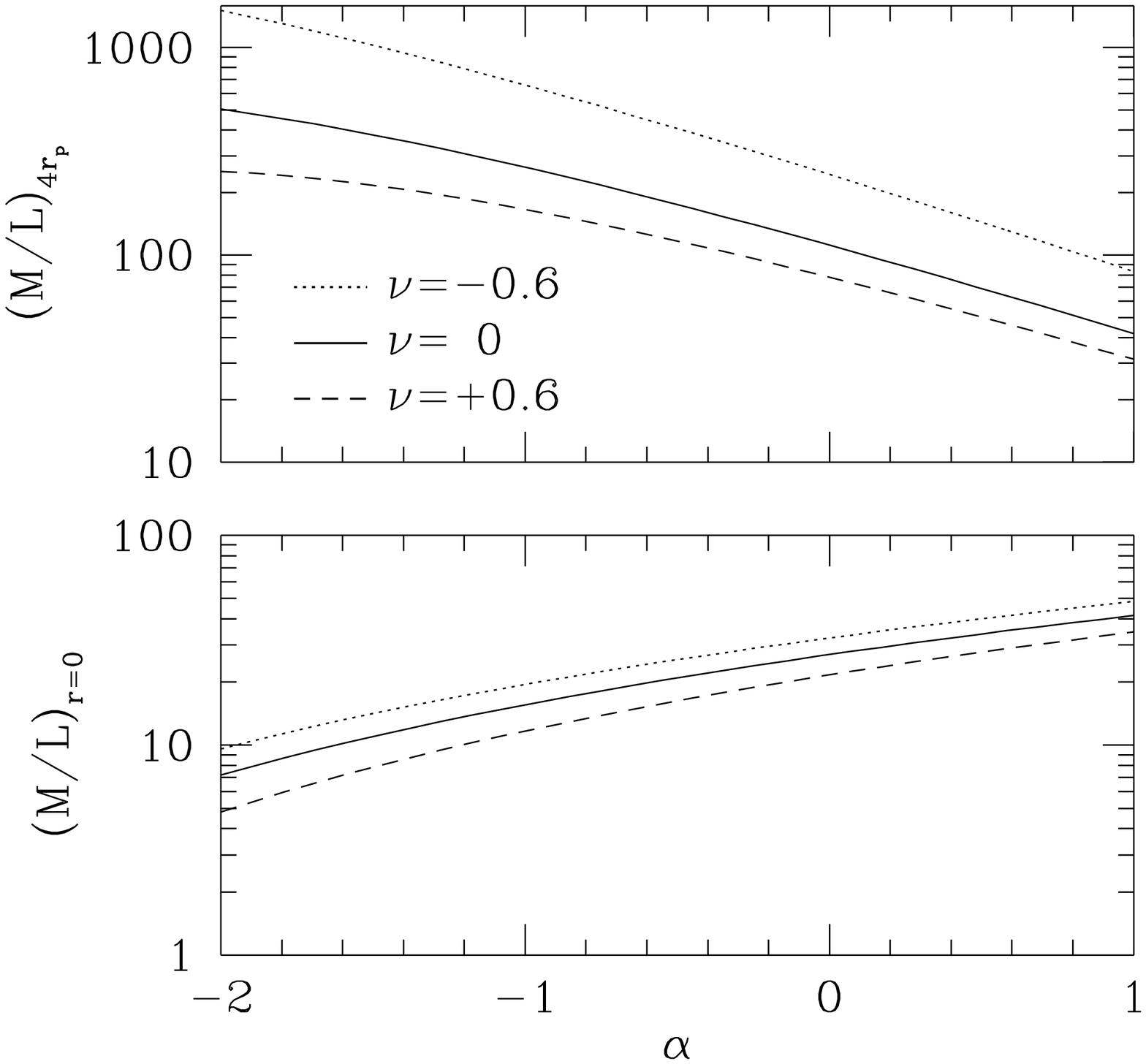,width=1.2\hssize}}
\smallskip\noindent
\caption{{\bf Figure \masstolight.} Variation of mass-to-light ratio (in
units of $M_\odot/L_\odot$) as a function of $\alpha$ for three
different $\nu$ values.  The parameter $\nu$ measures the anisotropy
of the velocity distribution. Models with $\nu = 0$ are everywhere
isotropic, while models with positive (negative) $\nu$ become
increasingly radial (tangential) at large radii (see section 2.3 for
discussion). The top panel shows the total $M/L$ within $4 r_0$ ($3
r_0 \approx$ Draco King tidal radius; Irwin \& Hatzidimitriou 1995).
The bottom panel shows the central $M/L$.}
\fignew
\endfigure

\section{Dwarf Spheroidal Models}
\subsection{Potential and Density}
In the past, dSphs have often been fitted using isotropic, spherical,
single component King (1962) models (e.g., Hodge 1966). These are
flexible and convenient, but they do come at the cost of very strong
assumptions -- namely that mass follows light and that all the stars
have a Maxwellian distribution of velocities out to the tidal radius.
Anisotropic King models relax the latter assumption, but still assume
that the distributions of mass and light are identical. Representation
of the dark matter and stars as two components of a multi-mass King
model is not physically appropriate as such models assume energy
equipartition among the different mass classes, which is certainly not
the case in a collisionless system such as a dSph. Noting this caveat,
Pryor and Kormendy (1990) applied such a model to data on the light
distributions and central velocity dispersions of the Draco and Ursa
Minor dSphs and found that models in which the luminous and dark
matter had similar distributions were favoured over models with more
extended dark matter distributions. However, the assumed coupling
between the dark and luminous matter means that changing the velocity
anisotropy of these models affects the distribution of the luminous
matter. In this paper, instead, we build a family of fully consistent
distribution functions for the stars in a dSph, where we assume that
the stars are tracer particles, moving in the underlying dark matter
potential. This allows us to probe the mass-anisotropy degeneracy
discussed above.

Plummer's (1911) model was originally developed to fit the light
distribution of the globular clusters, but a much better application
is to fit the light distribution of the dSphs (Lake 1990).  For
example, Figure~\figstarcount\ shows the best fitting Plummer profile
as compared to the background-subtracted star count data of two
Galactic dSphs, Draco and Sculptor, as given in Irwin \&
Hatzidimitriou (1995).  With the exception of the Sagittarius and Ursa
Minor dSphs, all the remaining seven Galactic dSphs are roughly
spherical, with ellipticities lying between 0.13 (Leo~II) and 0.35
(Sextans). So, the assumption of spherical symmetry is reasonable. We
envisage our models as being particularly useful for Draco and
Sculptor, because they are nearby and a wealth of kinematical data is
either already available or will become so over the next few years.

Accordingly, let us take the luminosity density of the dSph as 
$$
\eqnam{\plummer_rho_eq}
\rho (r)={\rho_0 \over \left[ 1+\left(r/r_0\right)^2\right]^{5/2}},
\eqno\new$$
where $\rho_0$ is determined by the total observed luminosity.  The
surface brightness of the dSph is
$$\Sigma (R) = {4\over 3} {\rho_0 r_0 \over \left[
1+\left(R/r_0\right)^2\right]^2},\eqno\new$$
where $R$ is the projected radius. The physical meaning of $r_0$ is
that it is the radius of the cylinder that contains half the light;
henceforth $r_0$ is set to unity.

Our aims are to assess the severity of the degeneracy between velocity
anisotropy and mass in the dSphs and to investigate what radial
velocity surveys may teach us. We need a flexible family of dSph
models with differing dark matter distributions and velocity
anisotropies, but fitting the same star count profiles on the sky. So,
we assume that the potential of the system has the form
$$
\eqnam{\model_psi_eq}
\psi(r)=\cases{{\displaystyle \psi_0 \over \displaystyle
\left[1+r^2 \right]^{\alpha/2}}
&if $\alpha \neq0$ ,\cr
\null &\null \cr
-{\displaystyle v_0^2 \over \displaystyle 2}\log \left[1+ r^2\right] &
 if $\alpha =0$,}
\eqno\new
$$
where $-2 \le \alpha \le 1$.  Setting $\psi_0 = v_0^2 / \alpha$, the
circular velocity curve is
$$\vcirc^2 = v_0^2 { r^2 \over \left[1+r^2\right]^{1 +
\alpha/2}}.\eqno\new$$
In other words, the circular velocity curve falls off asymptotically
like $r^{-\alpha/2}$.  This dark matter potential nicely spans the
range of dark matter density distributions which we wish to probe:
$\alpha=1$ corresponds to a mass-follows-light Plummer potential and a
Keplerian fall-off at large radii; $\alpha=0$ yields an asymptotically
flat rotation curve; and $\alpha=-2$ gives a harmonic oscillator
potential corresponding to the central regions of an extended
dark-matter halo.  Note that, as the parameter $\alpha$ decreases, the
dSph becomes more and more dark matter dominated. 

Figure~\masstolight\ shows the variation of the central and average
mass-to-light ratio as a function of the model parameters $\alpha$
(which controls the amount of dark matter present) and $\nu$ (which
measures the anisotropy of the velocity distribution -- see section
2.3). It is important to bear in mind that these mass-to-light ratios
correspond to models with the same star count density and the same
central velocity dispersion. The figure provides a telling indication
of the severity of the degeneracy between anisotropy and mass, since
the mass-to-light ratio varies by an order of magnitude as we scan the
models. It is this degeneracy we wish to break with kinematic
measurements.

\subsection{Distribution Functions}

We now build the distributions of velocities that supports the dSph
stellar density (\plummer_rho_eq) in the dark matter potential
(\model_psi_eq). As is well known, once the velocity distribution is
permitted to be anisotropic, there are many possible ways of building
a given density from stellar orbits (see Binney \& Tremaine 1987,
chap. 4).  The even part of the distribution function (DF) is
determined by the stellar density law, the odd part by the stellar
streaming or rotation law. There is no evidence for rotation in the
Draco or Sculptor dSphs, so we calculate only the even part of the DF.

\subsubsection{Isotropic DFs}

According to Jeans' (1915) theorem, the DF of the stars in a potential
is a function entirely of the isolating integrals of motion.
Isotropic models have DFs that depend only on the binding energy
$E$. The isotropic DFs are found from Eddington's (1916) inversion
formula as:
$$
\eqnam{\isodf}
F(E) = \cases{ C_{\alpha,0} |E|^{5/\alpha - 3/2} & if $\alpha \neq 0$, \cr
\null & \null \cr
C_{0,0} \exp\left( 5E/v_0^2 \right) & if $\alpha =0$,}\eqno\new$$
where the numerical constants $C_{\alpha,0}$ are given in
Table~\constants. These are very simple results -- compare, for example, the
much more complicated isotropic DFs of the spherical isochrone
(written out in Binney \& Tremaine 1987) or the Hernquist (1990)
sphere. The only other spherical models with comparably simple DFs are
the power-law spheres (Evans 1993, 1994).

\subsubsection{Anisotropic DFs: $\alpha >0$}

Anisotropic DFs of spherical models depend on both the energy $E$ and
the norm of the angular momentum $L$.  The DFs contain an additional
parameter $\gamma$, which we shall see shortly controls the
anisotropy. We begin by writing the Plummer density (\plummer_rho_eq)
in terms of the spherical polar radius $r$ and the potential
$\psi(r)$. There are many ways to do this, and each one corresponds to
a different anisotropic DF. We choose the density partition
$$\rho = \rho_0 \Bigl( {\psi\over \psi_0} \Bigr)^{5/\alpha -
\gamma/\alpha} (1+ r^2)^{-\gamma/2}.\eqno\new$$
For the case of a falling rotation curve ($\alpha > 0$), we can
exploit the results in Appendix A to obtain the DF as
$$
\eqnam{\anisodfa}
F(E, L^2) = C_{\alpha, \gamma} |E|^{(5 - \gamma)/\alpha - 3/2}
H(E,L^2).\eqno\new$$
When $L^2\le2|E|$, then
$$
\eqnam{\firsth}
H(E,L^2) = {}_2F_1 \Bigl(\frac{\gamma}{2}, \frac{3}{2}
   \!+\!\frac{\gamma-5}{\alpha},1 ; \frac{L^2}{2E} \Bigr),\eqno\new$$
whereas when $L^2\ge 2|E|$, then
$$
\eqnam{\secondh}
H(E,L^2) = \Bigl| {2E\over L^2} \Bigr|^{\frac{\gamma}{2}} {}_2F_1
   \Bigl(\frac{\gamma}{2}, \frac{\gamma}{2},
   \frac{\gamma-1}{2}\!+\!\frac{5-\gamma}{\alpha}; \frac{2E}{L^2}
   \Bigr).\eqno\new$$
Here, ${}_2F_1$ is the hypergeometric function of Gauss, while the
numerical constants $C_{\alpha,\gamma}$ are given in Table~\constants.
This result generalises an earlier calculation by Dejonghe (1987),
which is restricted to the case where mass follows light ($\alpha
=1$). Note that (\secondh) is not the analytic continuation of
(\firsth) beyond the unit circle $\frac{L^2}{2E}=1$.

It is worth remarking that when $\gamma = -2n$, the hypergeometric
function reduces to a polynomial of order $n$ in $\frac{L^2}{2E}$,
and then the DFs (\anisodfa) are entirely elementary. For example,
when $\gamma = -2$, we obtain 
$$
\eqnam{\earliercase}
F(E,L^2) = C_{\alpha,-2} |E|^{7/\alpha-3/2}
\Bigl[ 1 - (\frac{3}{2}-\frac{7}{\alpha})\frac{L^2}{2|E|}\Bigr].
\eqno\new$$

\subsubsection{Anisotropic DFs: $\alpha <0$}

For the case of a rising rotation curve ($\alpha < 0$), we use the
density partition
$$\rho = \rho_0 \Bigl( {-\psi\over -\psi_0} \Bigr)^{5/\alpha -
\gamma/\alpha} (1+ r^2)^{-\gamma/2}.\eqno\new$$
We develop the necessary formula for the anisotropic DF corresponding
to this partition in Appendix A. The result is
$$
\eqnam{\anisodfb}
F(E, L^2) = C_{\alpha, \gamma} |E|^{(5\!-\!\gamma)/\alpha\!-\!3/2}
G(E,L^2).\eqno\new$$
When $L^2 \le 2|E|$, the function $G(E,L^2)$ takes the form;
$$
\eqnam{\firstg}
G(E,L^2) = {}_2F_1 \Bigl(\frac{\gamma}{2}, \frac{3}{2}
 \!+\!\frac{\gamma-5}{\alpha},1 ; \frac{L^2}{2E} \Bigr),\eqno\new$$
whereas when $L^2 \ge 2 |E|$, we have;
$$
\eqnam{\secondg}
\eqalign{G(E,L^2)&= 
   \Bigl| \frac{2E}{L^2} \Bigr|^{\frac{\gamma}{2}} 
{}_2F_1 \Bigl(\frac{\gamma}{2}, \frac{\gamma}{2},
   \frac{\gamma\!-\!1}{2}\!+\!\frac{5\!-\!\gamma}{\alpha}; 
   \frac{2E}{L^2} \Bigr)\!+\!\Bigl| \frac {2E}{L^2}
   \Bigr|^{\frac{3}{2}\!+\!\frac{\gamma\!-\!5}{\alpha}} \cr
&\times{\Gamma(\lb)\Gamma(\frac{\gamma}{2}\!-\!\mlb)
\Gamma(1\!-\!\frac{\gamma}{2}) \over
   \Gamma(\frac{\gamma}{2})\Gamma(\cmb)\Gamma(\lb\!-\!\frac{\gamma}{2})}\cr 
&\times 
 {}_2F_1\Bigl(\frac{3}{2} 
   \!+\!\frac{\gamma\!-\!5}{\alpha}, \frac{3}{2}\!+\!
   \frac{\gamma\!-\!5}{\alpha},\frac{5\!-\!\gamma}{2}\!+\!
   \frac{\gamma\!-\!5}{\alpha}; 
   \frac{2E}{L^2}\Bigr)\cr}\eqno\new$$
This time, (\secondg) is the analytic continuation of (\firstg) beyond
the unit circle (e.g., eq [15.3.7] of Abramowitz \& Stegun 1970).

Again, it is worth remarking that when $\gamma = -2n$, the expression
reduces to a polynomial of order $n$ in $\frac{L^2}{2E}$, and then
the DFs (\anisodfb) are entirely elementary. For example, when $\gamma
= -2$, we obtain
$$F(E,L^2) = C_{\alpha,-2} |E|^{7/\alpha-3/2}\Bigl[ 1 - (\frac{3}{2}
-\frac{7}{\alpha})\frac{L^2}{2|E|}\Bigr],\eqno\new$$
which is the same as above (\earliercase).

\subsubsection{Anisotropic DFs: $\alpha =0$}

The case of a flat rotation curve ($\alpha =0$) has a different
form again. We write the Plummer density as
$$\rho = \rho_0 \exp \left( (5-\gamma)\psi/v_0^2 \right)
(1+r^2)^{-\gamma/2}.\eqno\new$$
The DF corresponding to this density partition is derived in Appendix
A. We find:
$$
\eqnam{\flatcase}
F(E,L^2) = C_{0,\gamma} \exp \left( (5-\gamma)E/v_0^2\right) \Phi
\Bigl( \frac{\gamma}{2},1, \frac{(\gamma-5)L^2}{2v_0^2} \Bigr),\eqno\new
$$
where $\Phi$ denotes the degenerate hypergeometric function and the
constant $C_{0,\gamma}$ is given in Table~\constants. When $\gamma =
-2n$, the function $\Phi$ reduces to the Laguerre polynomial $L_n$
(see eq. [8.972.1] of Gradshteyn \& Ryzhik 1978):
$$
\eqnam{\flatcaselag}
\eqalign{F(E,L^2) =& \rho_0 \Bigl( { 2n+5\over 2\pi v_0^2} \Bigr)^{3\over2}
L_n \Bigl[ {-(2n+5) L^2 \over 2 v_0^2} \Bigr] \cr
&\qquad\qquad \times \exp\Bigl( (2n+5) E/ v_0^2 \Bigr).\cr}\eqno\new
$$
The DFs (\flatcase) are then entirely composed of elementary
functions, namely the product of a polynomial of the angular momentum
and the exponential of the energy.  It is worth writing out the lowest
member to emphasise its simplicity. When $\gamma = -2$, we have:
$$F(E, L^2) = \rho_0 \Bigl({7 \over 2\pi v_0^2}\Bigr)^{3\over2} \Bigl[
1 + {7L^2 \over 2 v_0^2} \Bigr] \exp\Bigl(7E/v_0^2\Bigr).\eqno\new$$

\subsubsection{Practical Evaluation of the DFs}

The DFs (\anisodfa), (\anisodfb) and (\flatcase) all depend on the
hypergeometric function in one way or another.  Numerical evaluation
of the hypergeometric function can be difficult.  In fact, none of the
algorithms for evaluating the hypergeometric function $_2F_1(a,b,c;z)$
presented in the standard reference books on numerical methods (e.g.,
Press et al. 1992) is applicable for all values of the parameters
$a,b$ and $c$. This is because very different strategies are required
depending on the magnitude and signs of the parameters.  Our
computational algorithm is presented in Appendix B. Note that in our
applications to calculations of the likelihoods in Section 4, it is
important that the DFs are calculated extremely accurately. The
obvious way of checking our numerical algorithm is to investigate
whether the integration of the DF over velocity space yields the
Plummer density. Typically, we find that the density is recovered to
better than one part in $10^6$.

\subsection{Intrinsic Moments}

Although the DFs (\anisodfa), (\anisodfb) and (\flatcase) have
differing forms according to whether the rotation curve is falling
($\alpha > 0$), flat ($\alpha =0$) or rising ($\alpha <0$),
nonetheless all physical quantities like the moments vary smoothly
and continuously with the rotation curve index $\alpha$.

The intrinsic velocity second moments are
$$
\langle v_r^2 \rangle = {\ds v_0^2 \over \ds (\alpha\!+\!5\!-\!\gamma)}
{\ds 1 \over \ds (1\!+\!r^2)^{\alpha/2}},\eqno\new
$$
$$\langle v_\theta^2 \rangle = \langle v_\phi^2 \rangle =
{\ds v_0^2 \over \ds (\alpha\!+\!5\!-\!\gamma)}
{\ds 1 \over \ds (1\!+\!r^2)^{\alpha/2}}
\left[1-{\ds \gamma \over \ds 2}{\ds r^2 
\over \ds 1\!+\!r^2}\right].\eqno\new
$$
The second moments converge and are positive definite provided $\gamma
< \min (\alpha\!+\!5,2)$.  

In terms of Binney's (1981) anisotropy parameter $\alpha$, the radial
and tangential velocity dispersion $\langle v_r^2 \rangle$ and
$\langle v_\theta^2 \rangle$ vary as
$$
\beta = 1 - {\langle v_\theta^2 \rangle \over \langle v_r^2 \rangle}
= {\gamma \over 2}  {r^2\over 1\!+\!r^2}.\eqno\new
$$
If $\gamma =0$, the velocity dispersions are isotropic.  Irrespective
of $\gamma$, the central regions of the model are always isotropic. At
large radii, the anisotropy becomes constant (c.f., H\'enon 1973;
Wilkinson \& Evans 1999). If $\gamma <0$, then the dispersion tensor
becomes tangential with increasing radius; if $\gamma >0$, it becomes
radial. The limit $\gamma \rightarrow -\infty$ is the circular orbit
model, while the limit $\gamma \rightarrow 2$ is built from radial
orbits alone in the outer parts.

It is also helpful to define the quantity $\nu$ which is related to
$\gamma$ by
$$
\nu = \log_{10} \left({2 \over 2-\gamma}\right)
$$
This quantity runs from $-\infty$ (tangential velocity distribution at
large radii) to $+\infty$ (radial velocity distribution at large
radii). The velocity distribution is everywhere isotropic for $\nu =
0$. This definition is helpful because the ranges of tangential and
radial anisotropy are symmetric about $\nu =0$. 

The intrinsic fourth moments are useful for diagnosing deviations from
pure Gaussianity and so we briefly list the results here:
$$
\langle v_r^4 \rangle = {\ds v_0^4 \over \ds (\alpha\!+\!5\!-\!\gamma)
(2 \alpha\!+\!5\!-\!\gamma) } {\ds 1 \over \ds (1\!+\!r^2)^{\alpha}},
\eqno\new
$$
$$\eqalign{\langle v_r^2 v_\theta^2 \rangle = \langle v_r^2 v_\phi^2 \rangle 
= {\ds v_0^4 \over \ds (\alpha\!+\!5\!-\!\gamma) 
(2 \alpha\!+\!5\!-\!\gamma)}  {\ds 1 \over \ds
(1\!+\!r^2)^{\alpha}}&\cr \times
\left[1-{\ds \gamma \over \ds 2}{\ds r^2 \over \ds 1\!+\!r^2}\right],&\cr}
\eqno\new
$$
$$\eqalign{\langle v_\theta^4 \rangle = \langle v_\phi^4 \rangle &=
 {\ds v_0^4 \over \ds 8(\alpha\!+\!5\!-\!\gamma) 
(2 \alpha\!+\!5\!-\!\gamma)}  {\ds 1 \over \ds
(1+r^2)^{\alpha}}\cr & \times
\left[(4\!-\!\gamma)(2\!-\!\gamma)\!+\!{\ds 2\gamma(2\!-\!\gamma) \over 
\ds 1\!+\!r^2} + {\ds \gamma (2\!+\!\gamma) \over (1\!+\!r^2)^2} \right].\cr}
\eqno\new$$
All the remaining components of this fourth rank tensor vanish.  The
fourth moments converge and are positive definite provided $\gamma <
\min (2\alpha\!+\!5,2)$.

\beginfigure*{\fignumber}
\fignam{\figprojmom}
\centerline{\psfig{figure=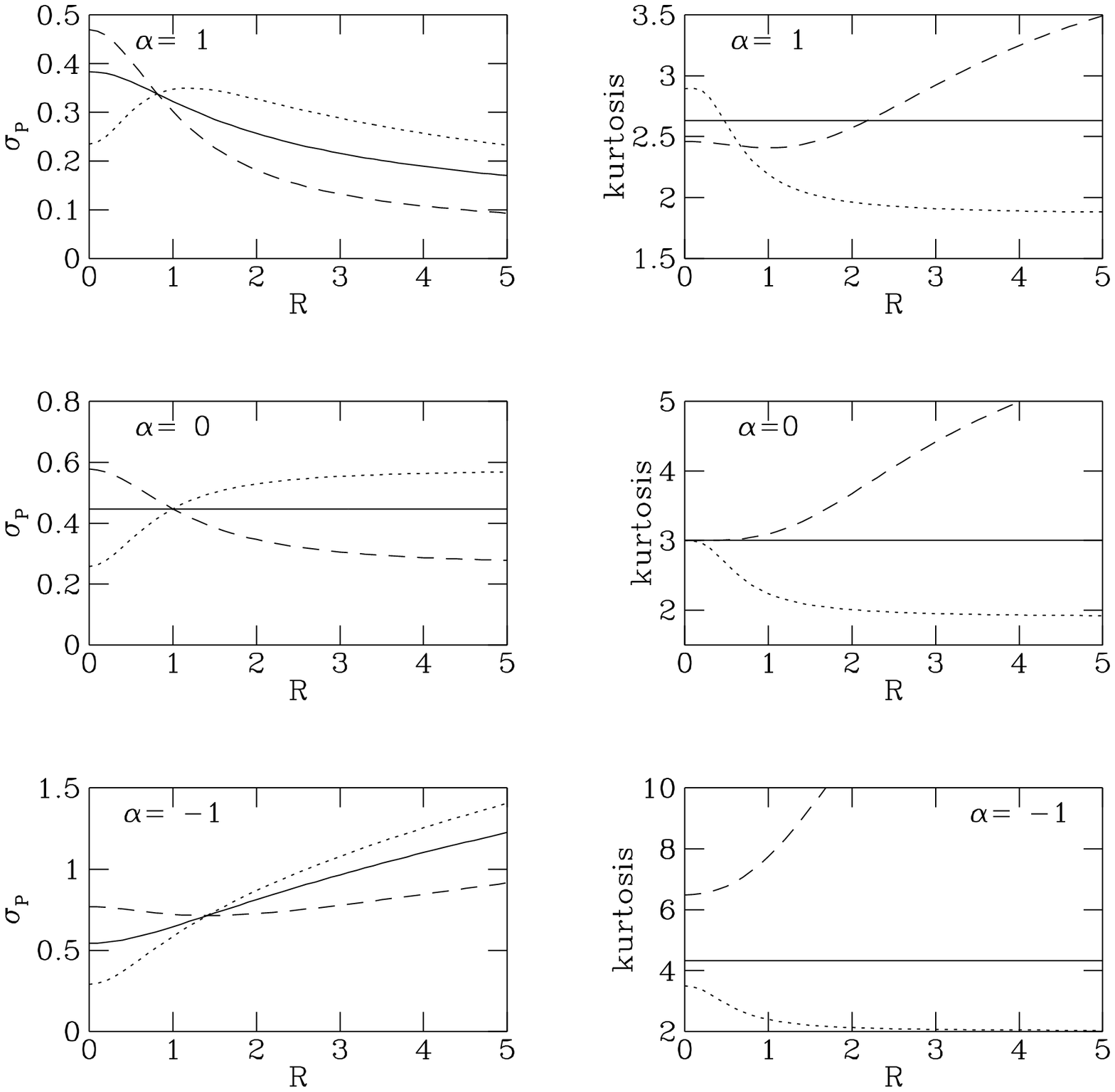,width=2.0\hssize}}
\smallskip\noindent
\caption{{\bf Figure \figprojmom.} This shows the variation of
the line of sight second moments (left panels) and kurtosis (right
panels) with radius R on the plane of the sky for models with
different dark matter content ($\alpha = 1,0,-1$ as labelled).  The
full curves refer to the isotropic model ($\nu=0$), the dashed
curve to a radially anisotropic model ($\nu = \infty$), the dotted
curves to a tangentially anisotropic model ($\nu=-0.8$).} 
\fignew
\endfigure
\beginfigure*{\fignumber}
\fignam{\LPfigone}
\centerline{\psfig{figure=lp_r_0.ps,width=1.5\hssize,angle=-90}}
\smallskip\noindent
\caption{{\bf Figure \LPfigone.} Normalised line profiles or
distributions of line of sight velocities $v_{\rm los}$ are shown for
models with different dark matter content ($\alpha = 1,0,-1$ as
labelled) and different anisotropy ($\nu = -0.6, 0, 0.6$). The line
profile is calculated at $R=0$, the center of the dSph in
projection. For comparison, the distributions of proper motions ($v_R$
and $v_\phi$) on the plane of the sky are also shown -- these profiles
are identical for $R=0$ and appear superposed as a dot-dashed line.}
\fignew
\endfigure
%
%
\beginfigure*{\fignumber}
\fignam{\LPfigthree}
\centerline{\psfig{figure=lp_r_4.ps,width=1.5\hssize,angle=-90}}
\smallskip\noindent
\caption{{\bf Figure \LPfigthree.} As Figure \LPfigone, but for
locations at projected position $R=4$.}
\fignew
\endfigure

\section{Observable Properties}

\noindent
The properties of dSphs when viewed in projection are of particular
interest to us, since these correspond to observable quantities.  In
this section, we present the distributions of the line of sight
velocities and the proper motions. All the moments of these
distributions are analytic.

\subsection{Line of Sight Moments}
To derive the line of sight velocity moments involves performing a
triple integration along the line of sight and over the two velocity
components perpendicular to the line of sight. For our dSph models,
these integrals are all analytic. The line of sight second moment is
given by
$$\sigmap^2 (R) = {\ds\sigma_0^2 \over \ds(1+R^2)^{\alpha/2}}
\left[1-{\gamma(4\!+\!\alpha)\over 2(5\!+\!\alpha)}{R^2
\over 1+R^2}\right]\eqno\new$$
where the central velocity dispersion is
$$\sigma^2_0 = {3\sqrt{\pi} v_0^2 \Gamma\left(2\!+\!\alpha/ 2\right)\over
4(\alpha\!+\!5\!-\!\gamma )\Gamma\left(5/2\!+\!\alpha / 2\right)}.
\eqno\new$$
When $\alpha =1$, this reduces to the result found by Dejonghe (1987)
for his self-consistent Plummer models. The left panels of
Figure~\figprojmom\ show the variation of the line of sight second
moment with projected radius $R$ for $\alpha = -1,0,1$ and $\gamma =
2,0,-10$ (or, equivalently, $\nu = \infty, 0, -0.8$).  For $\alpha >0$
and $\gamma \le -\alpha(5\!+\!\alpha)/ (4\!+\!\alpha)$, the curves
$\sigmap(R)$ show a maximum at
$$
R^2 = {2(\alpha (5\!+\!\alpha) + \gamma(4\!+\!\alpha)\over
\alpha( \gamma(4\!+\!\alpha) - 2(5\!+\!\alpha) )}.\eqno\new
$$
As Figure~\figprojmom\ indicates, the models have an unusual and
attractive feature. For a given $\alpha$, all the curves pass through
the same point
$$R^2 = {2 \over 2\!+\!\alpha}, \qquad \sigmap^2 = {3\sqrt{\pi}v_0^2\over 8}
{\Gamma ( 2\!+\!\alpha/2)\over \Gamma (7/2\!+\!\alpha/2)}\left({2\!+\!\alpha \over 4\!+\!\alpha}\right)^{\alpha/2},\eqno\new$$
irrespective of $\gamma$. This radius is the {\it changeover
radius}. Radially anisotropic dSphs ($\gamma,\nu >0$) have less projected
velocity dispersion $\sigmap$ at large radii than isotropic dSphs, and
more $\sigmap$ at small radii. This follows from simple geometrical
considerations. Radial orbits contribute significantly to the line of
sight motion in the inner parts, but much less so in the outer parts.
Conversely, tangentially anisotropic dSphs ($\gamma,\nu <0$) have more
projected velocity dispersion $\sigmap$ at large radii than isotropic
dSphs, and less $\sigmap$ at small radii. The changeover radius is the
radius at which the transition from the central to the asymptotic
properties occurs.

The line of sight fourth moment is useful for detecting deviations
from Gaussianity. It can be calculated from the intrinsic fourth
moments as:
$$
\eqalign{
\sigmap^4 (R) =& {\ds \sigma^4_0 \over \ds(1\!+\!R^2)^{\alpha}}
\Bigl[1- {2\gamma (2\!+\!\alpha) \over ( 5\!+\!2\alpha)}{R^2 \over
1\!+\!R^2}\cr &\qquad\qquad
+ { \gamma (2\!+\!\gamma)(2\!+\!\alpha)(3\!+\! \alpha)\over
2(5\!+\!2\alpha)(7\!+\!2\alpha) }{R^4\over(1\!+\!R^2)^2}\Bigr],\cr}
\eqno\new$$
with the central value given by
$$\sigma_0^{4} = {9\sqrt{\pi}v_0^4 \Gamma\left(\alpha\!+\!2\right)
\over 4\left(\alpha\!+\!5\!-\!\gamma\right)
\left(2 \alpha\!+\!5\!-\!\gamma \right)\Gamma\left({5/ 2}\!+\!\alpha\right)}.
\eqno\new$$
In fact, all the line of sight moments are analytic. The $n$th moment
$\sigmap^{2n}$ is
$$
\eqnam{\Momentn}
\sigmap^{2n}(R) = {\sigma_0^{2n}\over (1\!+\!R^2)^{n\alpha/2}}
{}_3{\rm F}_2\Bigl(-n,\frac{\gamma}{2},\frac{ 4+n\alpha}{2}; 
1, \frac{5 + n\alpha}{2}; \frac{R^2}{1+R^2}\Bigr),
\eqno\new$$
where $_3F_2$ is the generalised hypergeometric function (which always
reduces to a finite $n$th order polynomial). The central value is
$$\sigma_0^{2n} = 2^{n-2} {3\psi_0^n
\Gamma(n\!+\!{1/2})\Gamma({5/\alpha}\!-\! {\gamma /
\alpha}\!+\!1)\Gamma ( 2\!+\!n\alpha/2)
\over \Gamma({5 / \alpha}\!-\!{\gamma/ \alpha}\!+\!n\!+\!1) 
\Gamma (5/2\!+\!n\alpha/2)}.\eqno\new$$

Figure~\figprojmom\ shows the variation of the kurtosis $\kappa =
\sigma_{\rm p}^4/ (\sigma_{\rm p}^2)^2$ with projected radius $R$. The
kurtosis measures the extent to which the distribution is peaked. A
Gaussian distribution has a kurtosis of 3. The larger the kurtosis,
the broader is the distribution.  The isotropic models ($\gamma,\nu =
0$) always have a constant kurtosis given by
$$\kappa = {4\over\sqrt{\pi}}
{\Gamma(\alpha\!+\!2)\Gamma(\alpha/2\!+\!5/2)\Gamma(\alpha/2\!+\!7/2)\over
\Gamma(\alpha\!+\!7/2)\Gamma^2(\alpha/2\!+\!2)}.\eqno\new$$
In fact, the $\alpha=\gamma,\nu =0$ model has a constant kurtosis of 3,
indicating that the line profiles of this model are Gaussian.
Figure~\figprojmom\ illustrates the tendency for increasing kurtosis
as the dSph becomes more dark matter dominated. This stems
predominantly from the larger tails in the intrinsic velocity
distributions. As $\alpha$ diminishes, there are more and more high
velocity stars in the tails.

\subsection{The Line Profiles}

When comparing models to the discrete radial velocities, the main
quantity of interest is the line profile. This is the probability
distribution of the line of sight velocities at a given projected
radius.  Mathematically, the unnormalised line profile $L(v_\los,R)$
is the integral of the DF along the line of sight and over the
tangential components of velocity. Let us use ($R,\varphi$) as polar
coordinates on the plane of the sky and $z$ as the coordinate along
the line of sight.

When $\alpha \le0$, the line profile is:
$$L(v_\los,R) = \int_{-\infty}^\infty dz\int_{-\infty}^\infty dv_R
\int_{-\infty}^\infty dv_\varphi F(E,L^2),\eqno\new$$
whereas when $\alpha >0$, the line profile is:
$$\eqalign{L(v_\los,R) =& 
\int_{z_-}^{z_+} dz \int^{(2\psi\!-\!v_\los^2)^{1/2}}_
{-(2\psi\!-\!v_\los^2)^{1/2}} dv_R\cr
\times&\int^{(2\psi\!-\!v_\los^2\!-\!v_R^2)^{1/2}}_
{-(2\psi\!-\!v_\los^2\!-\!v_R^2)^{1/2}} dv_\varphi 
F(E,L^2).}\eqno\new$$
If $\alpha \le 0$, then the integration over the line of sight extends
indefinitely. If $\alpha >0$, the stars with velocities $v_\los$ are
seen at projected position $R$ only if $v_\los^2 < 2\psi (R,z)$. This
provides upper and lower limits $z_\pm$ to the line of sight
integral. Often it is useful to divide by the surface brightness
$\Sigma(R)$ and consider normalised line profiles $\ell (v_\los,R) =
L(v_\los,R) / \Sigma(R)$.

For the isotropic models, the line profiles can be reduced easily to
single quadratures. When $\alpha\neq 0$, the unnormalised line
profiles have the form
$$\eqnam{\generallp}
\eqalign{L(v_\los,R) =& \sqrt{{2 \over \pi}}{K_\alpha \rho_0 
\over |\psi_0|^{5/\alpha}}\cr
\times& \int_R^\infty \Bigl|\psi(r)\!-\!\frac{1}{2}v_\los^2 
\Bigr|^{5/\alpha\!-\!1/2}{r\,dr \over \sqrt{r^2-R^2}}.\cr}\eqno\new$$
where
$$K_\alpha \cases{
{\displaystyle \Gamma(5/\alpha\!+\!1)\over
\displaystyle \Gamma(5/\alpha\!+\!1/2)}
& if $\alpha <0$,\cr
\null & \null \cr
{\displaystyle \Gamma(1/2\!-\!5/\alpha)\over
\displaystyle \Gamma(-5/\alpha)}& if $\alpha
>0$.\cr}\eqno\new$$
If and only if the rotation curve is flat ($\alpha =0$) and the model
is isotropic ($\gamma,\nu =0$), then the normalised line profile is a pure
Gaussian
$$\ell(v_\los,R) = \Bigl({5\over 2 \pi v_0^2}\Bigr)^{1/2} 
\exp\Bigl( -{5v_\los^2 \over 2v_0^2} \Bigr).\eqno\new$$
We emphasise that the line profiles vary smoothly and continuously
through the seemingly singular case of $\alpha =0$. This may be
verified by taking the limit $\alpha \rightarrow 0$ of (\generallp).

Although presently unmeasurable with ground-based telescopes, the
internal proper motions of stars in the nearby dSphs are accessible to
future space missions (like the {\it Space Interferometry Mission} or
SIM). In Draco and Sculptor, the stars of interest are at magnitudes
$V \sim 19-20$ and the required proper motion accuracy corresponding
to $1-2\,{\rm km\,s }^{-1}$ in velocity is $3-6\,\mu{\rm as\,
yr}^{-1}$, which is just within the capabilities of SIM.  Hence, it is
also useful to calculate the one dimensional distributions of proper
motions $L(v_R,R)$ or $L(v_\varphi,R)$ by integrating along the line
of sight and over the other two transverse velocities. 

Figures \LPfigone\ and \LPfigthree\ present some examples of line
profiles and proper motion distributions obtained for a variety of
values of $R$, $\alpha$ and $\gamma$. These are calculated by direct
integration over the DFs. For the isotropic models ($\gamma,\nu =0$),
the line profile and the proper motion distributions clearly coincide.
Tangentially anisotropic clusters tend to show bimodal line profiles,
whereas radially anisotropic clusters tend to have narrower, peaked
line profiles. Hence, the trend as we move vertically downwards in the
panels is towards broader and more flat-topped profiles. The larger
the dark matter content, the greater and more distended the wings of
the line profiles. So, the trend as we move left to right in the
panels is towards less extended line profiles.

Note that, as we move outwards from the center of the dSph, the
differences between the line profiles become more
pronounced. Observationally speaking, it is easier to obtain radial
velocities for stars in the centre, but it is stars in the outer parts
of the dSph which are most useful for breaking the degeneracy between
mass and anisotropy.

\beginfigure*{\fignumber}
\fignam{\bincorrect}
\centerline{\psfig{figure=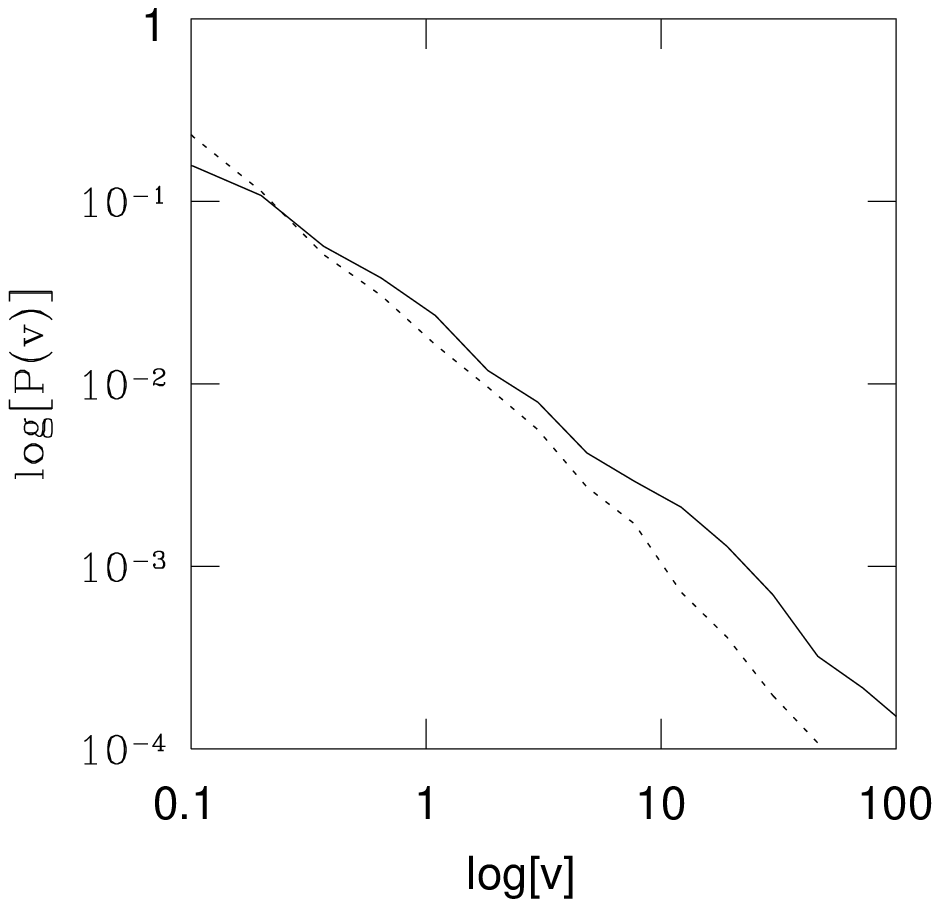,width=\hssize}}
\smallskip\noindent
\caption{{\bf Figure \bincorrect.} Distribution of binary velocities
as inferred from the binary distribution in the solar
neighbourhood. Dashed curve -- un-evolved distribution.  Solid curve
-- evolved distribution for giant star binaries. See text for
details.}
\fignew
\endfigure

\section{Methodology}

Previous studies of dSph dynamics have been impeded by the degeneracy
between anisotropy and mass. An increase in the line of sight velocity
dispersion at large radii may by due to either (1) the presence of
large amounts of mass at large radii, or (2) tangential anisotropy in
the velocity distribution. The large observed central radial velocity
dispersion is compatible with either a massive halo, and a low central
density; or no halo, and a large central density. In this section, we
test our ability to discern between these possibilities using (i)
large radial velocity surveys and (ii) radial velocities and proper
motions provided by SIM.

\subsection{The Likelihood Function}

The problem of comparing discrete radial velocity measurements to
galactic models has already received much attention (e.g., Little \&
Tremaine 1987, Kochanek 1996, Wilkinson \& Evans 1999, Evans \&
Wilkinson 2000). Suppose we are given radial velocities ${v_\los}_i$
(corrected for the galaxy mean velocity) of $N$ stars in a dSph at
projected positions $R_i$.  At each point, we can compute the
probability of observing the radial velocity data set
$$
\eqnam{\Prob_eq}
\eqalign{P( \{R_i,{v_\los}_i\}_{i=1\ldots N} | \alpha,\gamma) 
=& \prod_{i=1}^N \ell(R_i, {v_\los}_i; \alpha,\gamma).\cr}\eqno\new$$
So, we scan a grid of $\alpha,\gamma$, and at each point compute the
probability of observing the entire input data set.  Using Bayes'
theorem and assuming uniform prior probabilities in the model
parameters, then the most likely values of $\alpha$ and $\gamma$ are
given by maximising (\Prob_eq).  Confidence regions are obtained by
applying two-dimensional $\chi^2$ statistics to the likelihood (or the
logarithm of (\Prob_eq)).

This procedure requires repeated evaluation of the line profiles, and
direct integration over the DF is too slow and expensive to provide a
competitive algorithm. Accordingly, we only use brute force
integration to provide a look-up table of the line profiles in a grid
in the four-dimensional ($\alpha,\gamma, R,v$) space. The grid spacing
in $\alpha$, $R$ and $v$ is linear, whereas the spacing in $\gamma$ is
uniform in $\log (2-\gamma)$. We use cubic splines to interpolate in
the logarithm of the line profile between the grid points. Once the
look-up table has been built, this provides an extremely fast and
accurate way of calculating the line profiles.  Typically, the error
in the line profile (as inferred by integration over the line of sight
velocity to recover the surface density) is better than one part in
$10^3$. This is still sufficient to mislead our maximum likelihood
algorithm because the probability (\Prob_eq) is formed by the
multiplication of $N$ line profiles. To take account of this small
error, we re-normalise each line profile to unity after interpolation.

Before applying the likelihood algorithm, there are two further
corrections that must be applied.  First, the line profile is
convolved with a Gaussian of width $2\,{\rm km\,s}^{-1}$ to allow for
observational errors.  This is a typical error for data obtained on
the 4m class telescopes like the William Herschel Telescope with
multi-object spectrographs. Second, the line profile is adjusted for
the effect of binaries. This also involves convolution, this time with
the binary correction function $b(v)$.  By Monte Carlo sampling binary
orbits drawn from the binary distribution in the solar neighbourhood,
taking account of tidal circularisation of the orbits as described in
Paper II, we deduce the distribution of velocities $P_b(v)$ induced by
the binary motion. This is shown in Figure~\bincorrect. The binary
correction function is then given by
$$b(v) = f P_b(v) + (1-f) \delta(v),\eqno\new$$
where the constant $f$ is the binary fraction.

\subsection{Radial Velocity Surveys}

\beginfigure{\fignumber}
\fignam{\montefigone}
\centerline{\psfig{figure=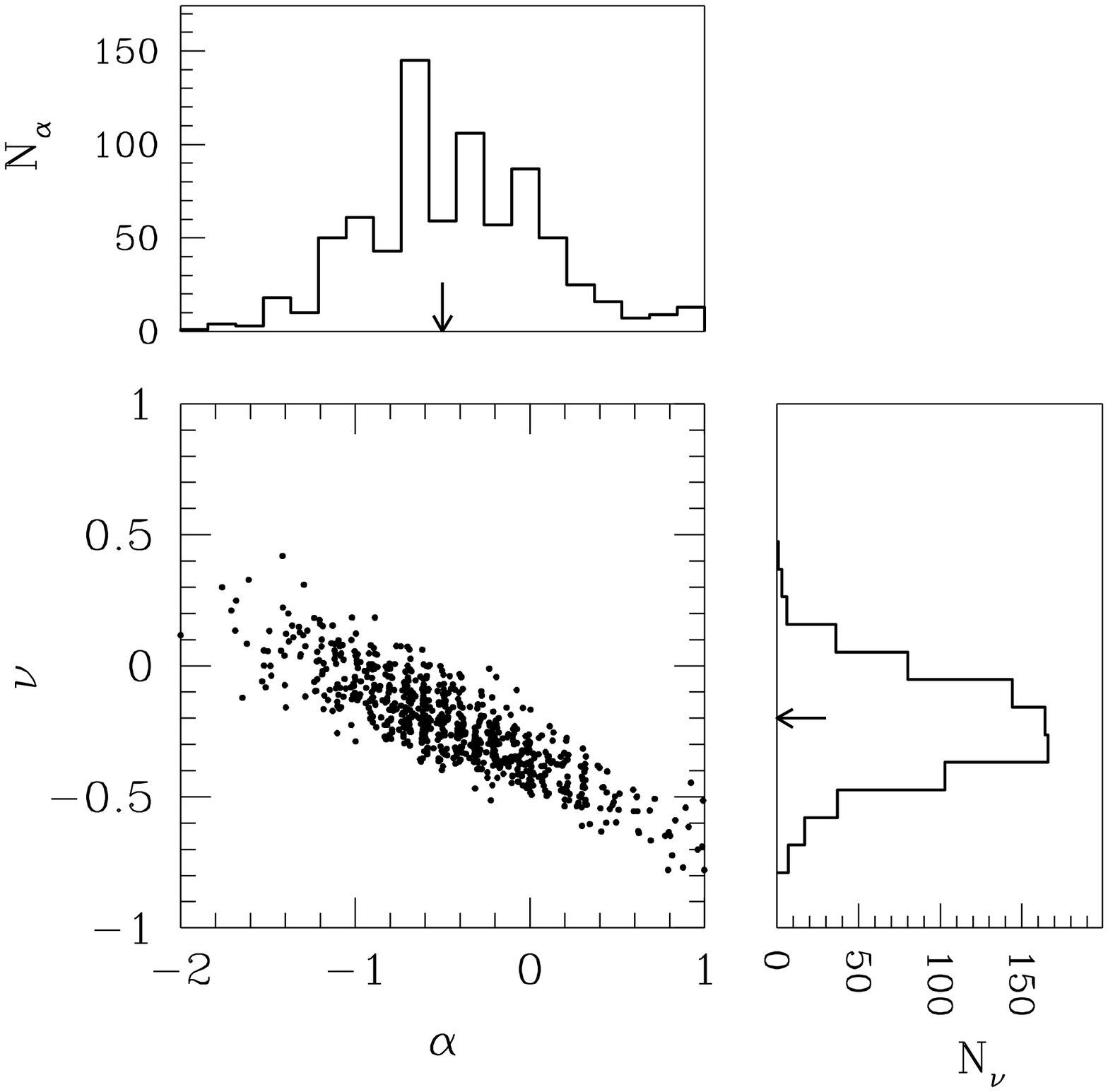,width=1.0\hssize}}
\smallskip\noindent
\caption{{\bf Figure \montefigone.} The recovery of the dark matter
parameter $\alpha$ and the anisotropy $\nu$ from simulated data sets of
160 radial velocities. Observational errors of $2$ km\,s$^{-1}$ are
assumed for the radial velocities and a binary fraction of $40\%$ is
assumed. The main panel shows where each data set falls in the
($\alpha,\nu$) plane. The side panels show the one-dimensional
histograms. The true values of $\alpha$ and $\nu$ are marked with
arrows.}
\fignew
\endfigure

\beginfigure{\fignumber}
\fignam{\montefigtwo}
\centerline{\psfig{figure=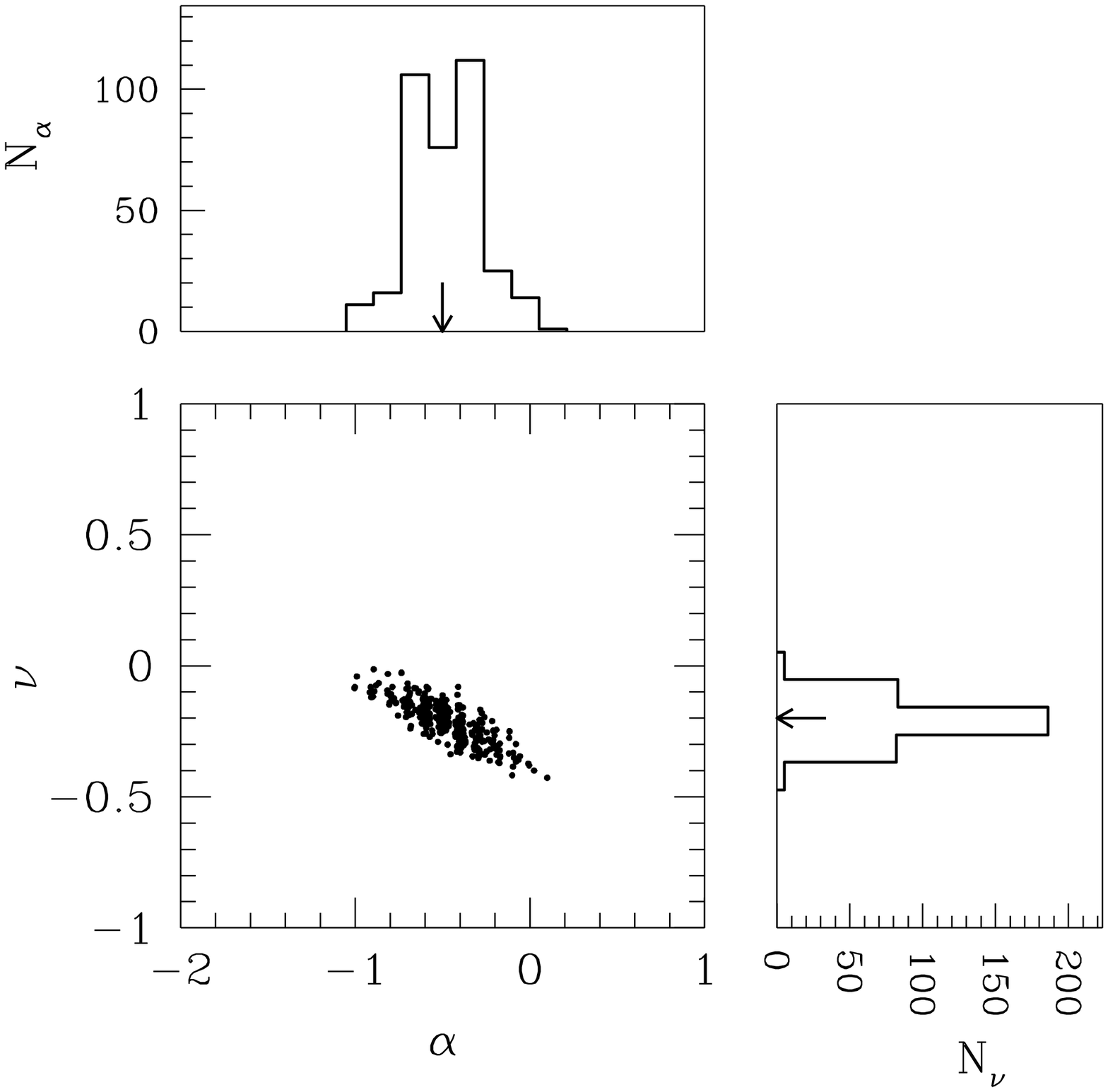,width=1.0\hssize}}
\smallskip\noindent
\caption{{\bf Figure \montefigtwo.} As Figure \montefigone, but
for simulated data sets of 1000 radial velocities each.}
\fignew
\endfigure

\beginfigure{\fignumber}
\fignam{\montefigthree}
\centerline{\psfig{figure=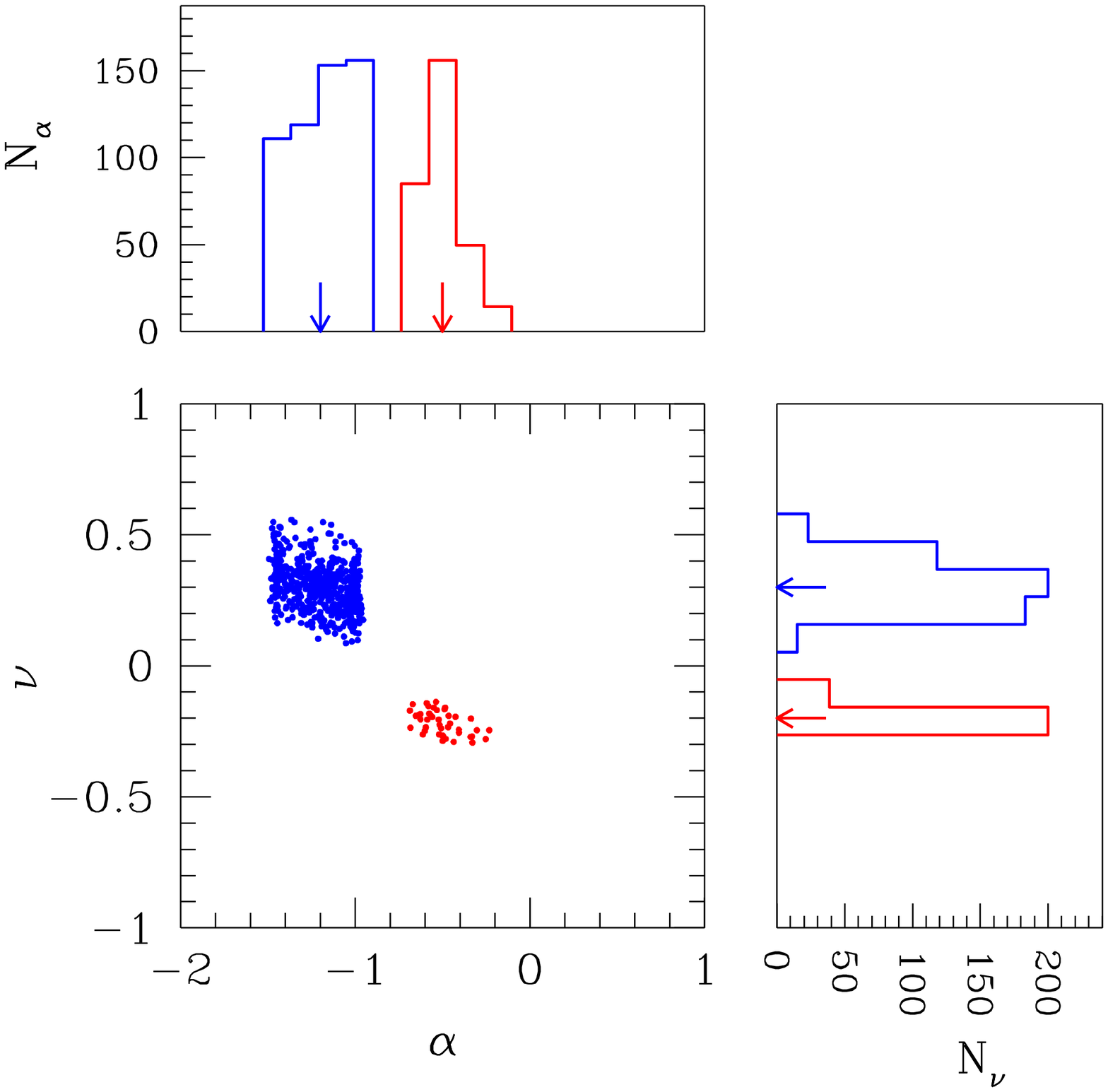,width=1.0\hssize}}
\smallskip\noindent
\caption{{\bf Figure \montefigthree.} The recovery of the dark matter
parameter $\alpha$ and the anisotropy $\nu$ for simulated data sets of
160 stars with ground-based radial velocities, together with proper
motions measured by SIM. The true values of $\alpha$ and $\nu$ are
marked with arrows.}
\fignew
\endfigure

\beginfigure{\fignumber}
\fignam{\montefigfour}
\centerline{\psfig{figure=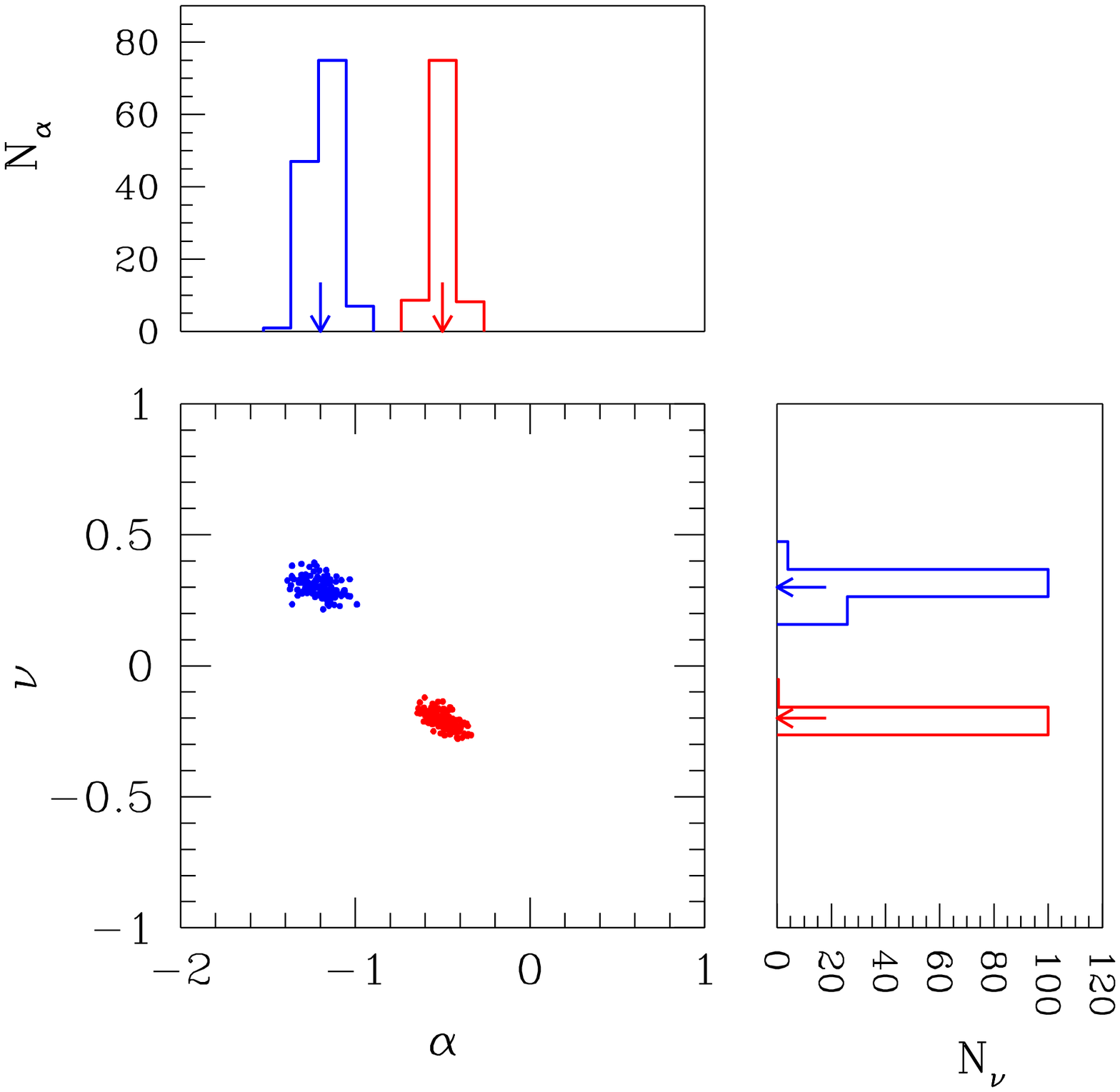,width=1.0\hssize}}
\smallskip\noindent
\caption{{\bf Figure \montefigfour.} As Figure \montefigthree, but
for simulated data sets of 1000 stars with radial velocities and proper
motions.}
\fignew
\endfigure

\beginfigure{\fignumber}
\fignam{\montefigfive}
\centerline{\psfig{figure=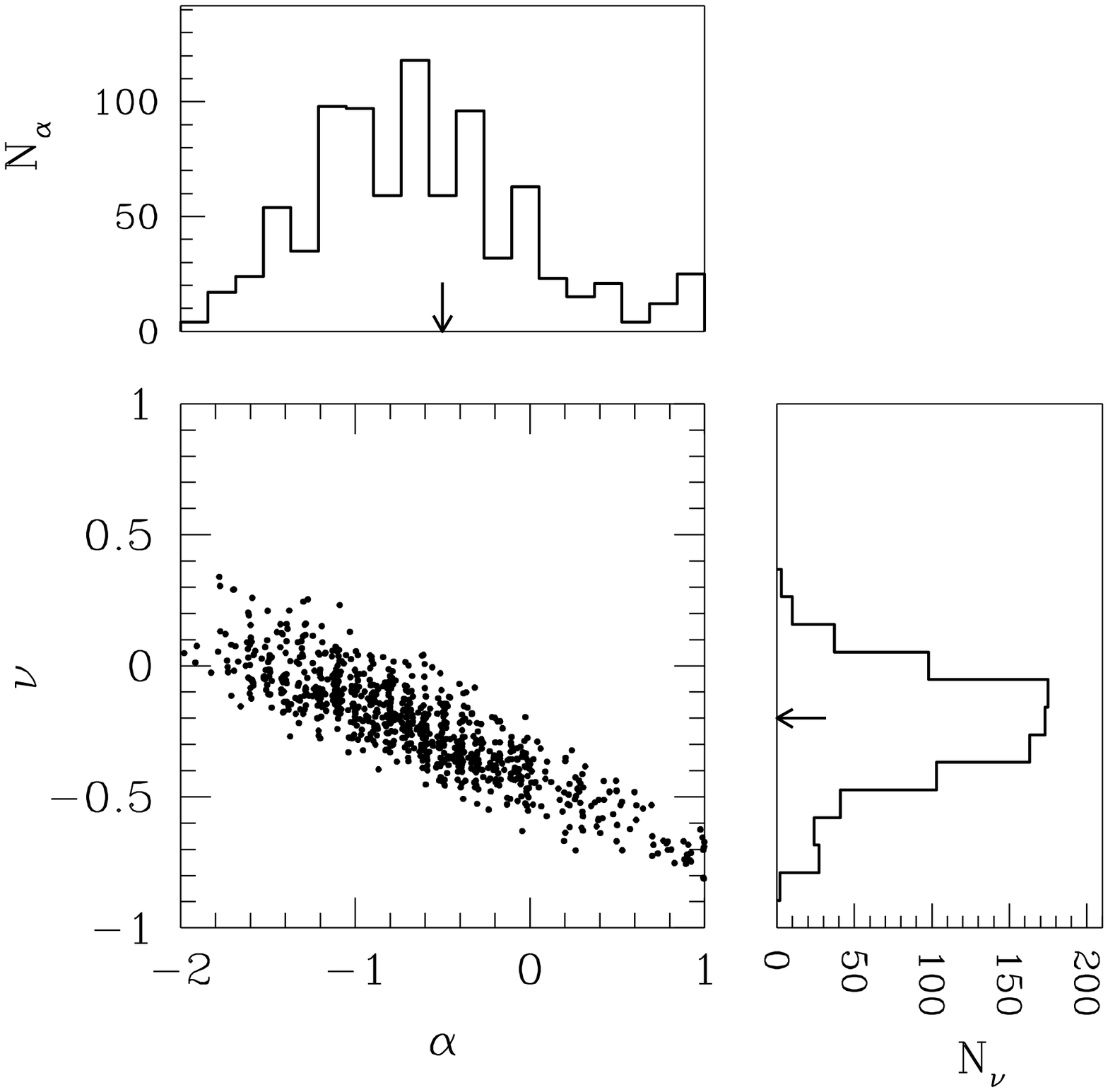,width=1.0\hssize}}
\smallskip\noindent
\caption{{\bf Figure \montefigfive.} The recovery of the dark matter
parameter $\alpha$ and the anisotropy $\nu$ for simulated data sets of
160 stars with radial velocities. Data produced with a binary fraction
of $80\%$ but analysed assuming a binary fraction of $40\%$.}
\fignew
\endfigure

\beginfigure{\fignumber}
\fignam{\figjaffe}
\centerline{\psfig{figure=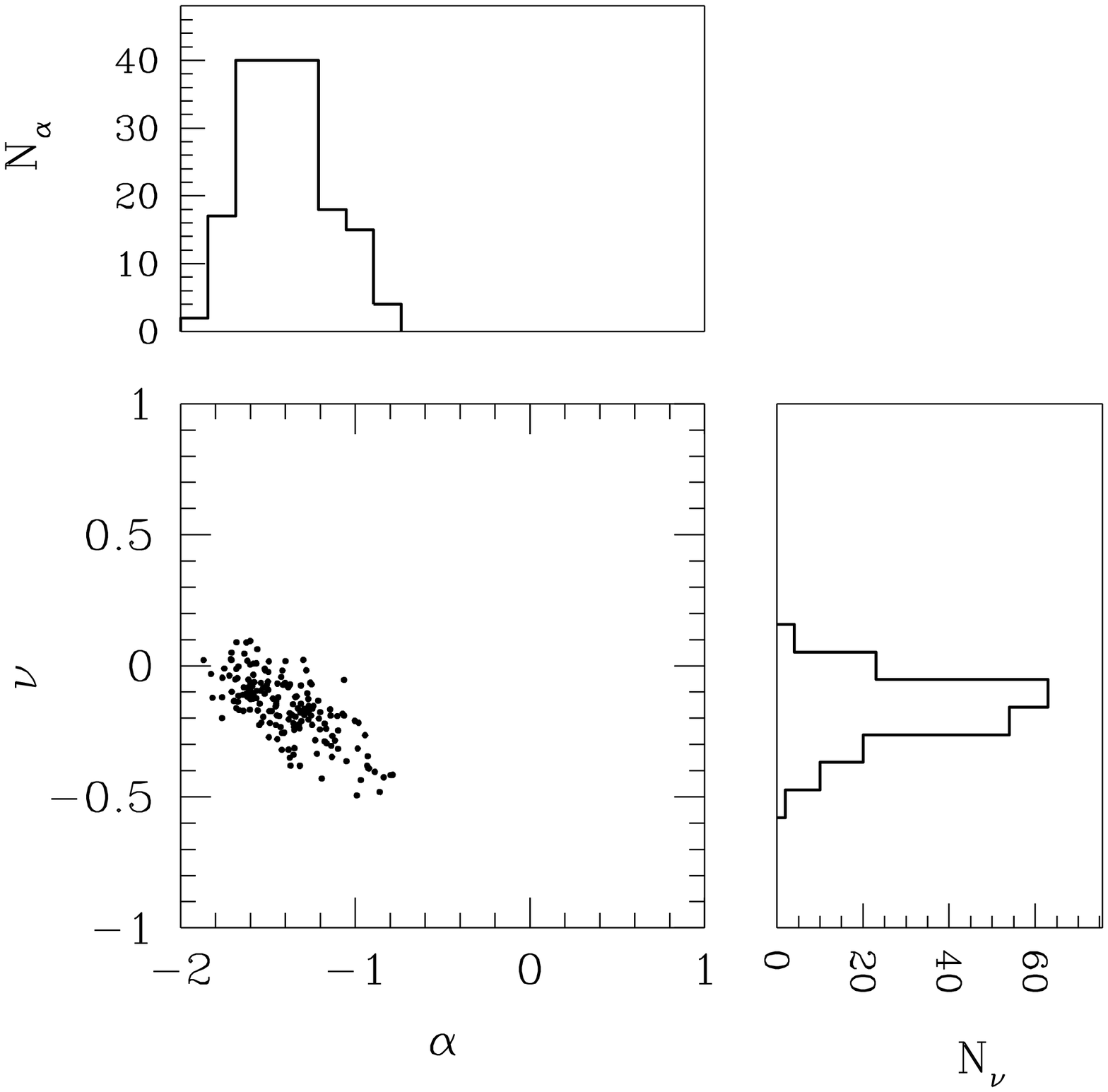,width=1.0\hssize}}
\smallskip\noindent
\caption{{\bf Figure \figjaffe.} The recovery of the dark matter
parameter $\alpha$ and the anisotropy $\nu$ for simulated data sets of
160 stars with radial velocities. Data generated from a model with a
Jaffe sphere representing the dark halo of the dSph. See text for
details.}
\fignew
\endfigure
%
%

Synthetic data sets are created by choosing phase space coordinates
$\{R_i,z_i,{v_R}_i,{v_\varphi}_i,{v_\los}_i\}_{i=1\ldots N}$ drawn
from the DF, and then discarding all but the projected positions and
the line of sight velocities. The velocities are contaminated by
measurement noise of amplitude $2\,{\rm km\,s}^{-1}$, and a fraction
$f$ of the stars are also assigned a binary velocity which is added to
their line of sight velocity. To reduce the computation time, rather
than randomly selecting the projected radius for each star, ten values
of the projected radius $R_i$ between 0.0 and 2.5 are used. Each data
set is then analysed and a most likely value of $\alpha$ and $\gamma$
(or, equivalently, $\nu$) is obtained.

Figure~\montefigone\ shows the results from the analysis of simulated
data sets of 160 radial velocities. This is the number of radial
velocities that are presently available for
Draco. Figure~\montefigtwo\ shows the results from analysis of
simulated data sets of 1000 radial velocities. This is an estimate of
the maximum number of radial velocities that may be available in Draco
with an observing program on a 4m class telescope. In each figure, the
scatter plot shows the joint distribution of $\alpha$ and $\nu$ values
obtained from the artificial data sets while the histograms show the
spread in the individual parameter values.

Even with a data set of the present size, Figure~\montefigone\ shows
that it is possible to put interesting constraints on the value of
$\alpha$.  The input model for Figure~\montefigone\ had an $\alpha$
value of $-0.5$ indicating a dark matter distribution intermediate
between that of a flat rotation curve halo and an extended harmonic
core. From the figure, the $2\sigma$ confidence interval on $\alpha$ is
$(-1.45,0.8)$. Thus data sets containing 160 radial velocities are
sufficient to rule out a mass-follows-light halo model ($\alpha = 1$)
at about the $2.5\sigma$ level. Increasing the size of the data set to
1000 stars allows much greater discrimination between models.

\subsection{Proper Motion Surveys}

Here, we consider the possible impact of next generation astrometric
satellites like SIM and its successors. Microarcsecond astrometry to
an accuracy $3-6\,\mu{\rm as\, yr}^{-1}$ is sufficient to allow the
internal proper motions of the brightest stars in Draco to be measured
to $1-2\,{\rm km\,s}^{-1}$.  Since all stars in Draco are at almost
the same distance, only the relative proper motions are required and
this significantly reduces the observing time. A further advantage is
that all the target stars are within a $\sim 1^\circ$ field of view,
allowing subdivision of the observations into a small number of
overlapping fields and hence reducing the instrumental systematic
contribution to the error. A satellite like SIM is ideal for these
measurements, as it allows very accurate differential measurements at
faint magnitudes.

To simulate this, we pick phase space coordinates
$\{x_i,y_i,{v_x}_i,{v_y}_i,{v_z}_i\}_{i=1\ldots N}$ drawn from the DF,
discard the geocentric radial position coordinate $z$ , and attempt to
recover $\alpha$ and $\nu$ using the Bayesian likelihood. As in the
radial velocity simulations described above, we contaminate the data
with measurement noise in each component of velocity. The correct
expression for the probabilities required to analyse these data is
then the 3-dimensional convolution of the DF with 3 Gaussians, each of
width $2\,{\rm km\,s}^{-1}$. However, as was noted earlier,
integration of the DF is computationally expensive, and we therefore
approximate the convolution by a weighted sum over the values of the
DF at the corners of a cube of side $\sqrt{2}\sigma$. This approach
gives sufficiently accurate results for the present paper -- in the
future, increases in computational speed will allow us to perform the
full 3-dimensional convolution. At present, we consider only the case
where no binaries are present. This is both computationally
convenient, and may be justified on the basis that proper motion
measurements will be based on multiple epoch observations which will
allow the identification of binaries in our sample.

Figure~\montefigthree\ shows the results from the analysis of
simulated data sets of 160 with both radial velocities and proper
motions measured for each star to an accuracy of $2\,{\rm
km\,s}^{-1}$. Figure~\montefigfour\ shows the results from analysis of
similar data sets containing 1000 stars. The narrow spread visible in
these histograms, particularly in the case of 1000 stars, emphasises
the value of having all components of the stellar velocities when
modelling the mass profile. The key feature of these results, as
illustrated in Figure~\montefigthree\ , is that even with only 160
proper motions it is possible to break unambiguously the degeneracy
between mass and anisotropy.

\subsection{Robustness}

In this section, we consider the robustness of our analysis to errors
in the estimation of the sample binary fraction and measurement
errors. As the same algorithm is used in both cases, we present
results only for the former case.  We also address the question of the
extent to which our analysis is model dependent.

Let us begin by considering data sets in which the binary fraction has
been severely underestimated. Data sets of 160 radial velocities are
generated as described earlier, assuming a binary fraction of $80\%$,
and measurement errors of $2\,{\rm km\,s}^{-1}$. However, when
analysing the data, we perform the convolution of the line profiles
assuming a binary fraction of only $40\%$. This discrepancy represents
an upper limit to the uncertainty in the binary fraction in a typical
data set of single-epoch radial velocities. The results of this
experiment are shown in Figure~\montefigfive.  Comparison of this
figure with Figure~\montefigone\ shows that while the distributions of
recovered $\alpha$ and $\gamma$ values are slightly broadened, the
overall effect of the mis-estimation of the binary fraction is not
serious. The uncertainty caused by the true size of observational
errors can be checked in a similar manner. We generate data sets of
160 radial velocities with a binary fraction of $40\%$ and measurement
errors of $4\,{\rm km\,s}^{-1}$. However, we convolve the line
profiles with an error Gaussian of width $2\,{\rm km\,s}^{-1}$
representing an over-optimistic estimation of the measurement
errors. We do not show a figure for this case, as the results are
predictable. Our algorithm is not confused and the results are
clustered around the correct values, albeit with a greater spread.

A more serious question is whether our maximum likelihood algorithm --
which assumes a parametrised fit for the density and potential -- is
robust, as any dSph undoubtedly deviates to a greater or lesser extent
from the model. To test this, we generate synthetic data from a very
different halo model and analyse them using our models. We retain the
assumption that the light in the dwarf follows a Plummer profile, but
we replace the halo with a Jaffe (1983) sphere. The halo potential is
then given by
$$\psi(r) = v_c^2 \log\left({r+r_{\rm J} \over r}\right)$$
where $v_c$ is the amplitude of the halo rotation curve at small
radii, and $r_{\rm J}$ is the scale radius of the halo. For these
simulations, we set $r_{\rm J} = 1$ and $v_c = 45.5$ km\,s$^{-1}$,
yielding a halo whose rotation curve falls off for $r > 1$ but for
which total mass enclosed within $r \sim 3$ is similar to that of an
$\alpha \sim -2$ (rising rotation curve) model from our family of
Plummer halo models. Following Wilkinson \& Evans (1999, Eq. (19)) we
build a DF (with constant velocity anisotropy $\beta$) of the form
$F(E,L) = L^{-2\beta} f(E)$. The value of $\beta$ is set to be the
mean value of the velocity anisotropy between $r=0$ and $r=3$ for a
model with $\nu = -0.2$ from our Plummer family. We analyse the data
using radial velocities alone.

Figure~\figjaffe\ shows the results obtained from data sets of 160
radial velocities. The analysis of the Jaffe model data sets returns
Plummer model parameter values in the region of $\alpha = -1.5$, $\nu
= -0.2$. Qualitatively, this result is very encouraging, as the
parameter estimates all lie in the region of the ($\alpha,\gamma$)
plane where the dark matter is more extended than the light in the
dSph. More quantitatively, the Jaffe data sets return approximately
the correct value of $\nu$. The mass enclosed within $r=3$, the region
probed by the generated data points, is also recovered to better than
a factor of two -- for the (input) Jaffe model we obtain $3.6\times
10^8$ M$_{\odot}$, while for the recovered Plummer halo we obtain
$2.2\times 10^8$ M$_{\odot}$. We conclude that the parameter values
returned by our method are reasonably robust, even for a small data
set, and that they do reflect underlying physical properties of the
data.

\section{Conclusions}

This paper has presented a flexible family of spherical models
suitable for representing dwarf spheroidal (dSph) galaxies. The models
have a Plummer profile, which is an excellent fit to the star count
data on the nearby dSphs. There are two free parameters. The first is
the dark matter parameter $\alpha$.  When $\alpha =1$, the mass
follows the light and the rotation curve is asymptotically Keplerian;
when $\alpha =0$, the dark matter is distributed in a cored isothermal
sphere and the rotation curve is flat; when $\alpha =-1$, the dSph is
enclosed within the harmonic core of a much larger dark matter
halo. The second parameter is the anisotropy parameter $\gamma$. When
$\gamma >0$, the models are radially anisotropic at large radii; when
$\gamma=0$, they are isotropic; when $\gamma <0$, they are
tangentially anisotropic at large radii.

These models offer considerable advantages over King models, which are
conventionally used for fitting dSphs (see e.g. Binney \& Tremaine
1987). The single component, isotropic King models often used for
modelling the dSphs assume that the mass follows the light and have an
isotropic stellar velocity distribution. The cost of such strong
assumptions is that they are unable to probe the degeneracy that
exists between dark matter mass and anisotropy.  A very attractive
property of our new models is that the intrinsic and projected moments
are simple and analytic. In particular, the line of sight velocity
dispersion is available as a function of projected radius for all
values of the dark matter and anisotropy parameters. The phase space
distribution function (DF) is more complicated -- as it depends on
transcendental functions -- but it is feasible to evaluate
numerically. Quadratures over the DF provide us with the distributions
of radial velocities (the line profiles) as well as the distributions
of proper motions. All these observable quantities are readily
available for our family of dSph models! Note that the line profiles
of all the models are similar in the centre, but begin to differ at
$\sim 2r_0$, emphasising the importance of gathering data at large
radii.

We have used our new models to assess what large samples of radial
velocities can tell us about dark matter in dSphs. All of the
information on the kinematics is contained in the DF. So, given large
samples of stars with projected positions and radial velocities, we
calculate the likelihood of observing the data set as a function of
the dark matter and anisotropy parameters.  The distribution of radial
velocities is corrected first for observational errors by convolution
with a Gaussian and second for the contamination by binaries by
convolution with a suitable correction function.

Simulated data sets of 160 and 1000 radial velocities are drawn from
the DF, contaminated with measurement noise and binary velocities, and
fed into the likelihood algorithm. A sample size of hundreds of
velocities is typical of the presently available data sets, whereas a
sample size of a thousand velocities is an estimate of the number
that may be accessible to a long-term observing program in the nearby
dSphs like Draco. Our simulations show that a sample size of 160 stars
is already sufficient to make interesting statements about the dark
matter distribution in a dwarf galaxy. In particular, using data drawn
from a model with a slightly rising rotation curve it is possible to
rule out mass-follows-light models at about the $2.5\sigma$ confidence
level. Larger radial velocity data sets allow much tighter constraints
to be placed on the model parameters.  Our machinery is set to work in
a companion paper (Kleyna et al. 2001) to interpret a newly acquired
data set for the Draco dSph.

The addition of proper motion data completely breaks the degeneracy
between mass and velocity anisotropy, even using only a hundred or so
proper motions. Obtaining such a data set will be feasible using the
SIM satellite. One strategy is to construct a sample of stars that are
uncontaminated by binaries before SIM flies, for example, by taking
second and third epoch data on ground-based telescopes and removing
all radial velocity variables.  SIM time is very expensive and it is
wasteful to spend it following the astrometric paths of binaries if
our aim is to measure the proper motions of single stars. In Draco and
Sculptor, the target stars are at magnitudes $V \sim 19$ and the
required proper motion accuracy corresponding to $1-2\,{\rm km\,s
}^{-1}$ in velocity is $15-30\,\mu{\rm as\,}$ over the mission
lifetime of 5 years.  The optimum strategy is to measure the positions
of the stars twice, once at the beginning and once at the end of the
mission. This will be ample to calculate the proper motions, provided
all binaries have been eliminated by our ongoing radial velocity
surveys. SIM takes $\sim 1$ hour to measure the position of $V \sim
19$ star to $\sim 20\,\mu{\rm as\,}$ (in one dimension), which is our
typical required accuracy. Hence, for each star, we require 4 hours of
SIM time (for two-dimensional measurements at the beginning and end of
the mission).  Our simulations have shown that samples of $\sim 100$
proper motions are ample to break the mass-anisotropy degeneracy. This
will take about 400 hours of SIM time, or roughly $1.7\%$ of the
mission lifetime.  In other words, {\it this is an extremely
competitive use of SIM time.}

Given the modest amount of SIM time required, our collaboration is
considering an ambitious program that acquires the proper motions of
roughly a hundred stars in each of Draco, Ursa Minor and Sextans.
This program will still only consume $5 \%$ of the mission lifetime.
Such a data set, together with the sophisticated modelling techniques
we have introduced in this paper, will make it possible to map the
dark matter distribution in these three dSph galaxies with
unprecedented accuracy. This important test of the nature of dark
matter on such small scales is uniquely possible with SIM.

\section*{Acknowledgments}
MIW and JK acknowledge financial support from PPARC. NWE thanks the
Royal Society for financial support and Andy Gould for a number of
insightful remarks about the Space Interferometry Mission.

\appendix

\section{Derivation of the Distribution Functions}

First, let us summarise Dejonghe's (1986) results, which apply to the
models with falling rotation curves ($\alpha >0$). Let us start with
the well-known result (Fricke 1951, Dejonghe 1987, Evans 1994) that the
density $\rho'$
$$\rho' = \psi^p r^{-\beta},\eqno\new$$
corresponds to the DF
$$f' = {\Gamma (p+1)\over \Gamma( 1- \beta/2)(p - 1/2 + \beta/2)}
{L^{-\beta}E^{p+\beta/2 - 3/2} \over (2\pi)^{3/2}
2^{-\beta/2}}.\eqno\new$$
These are sometimes called Fricke components.  From this simple
result, we can build up the DF $F$ corresponding to the density
$$\rho = \psi^p {r^{2a} \over (1+r^2)^{a+b}},\eqno\new$$
by a continuous superposition of the Fricke components. The result is
$$F(E,L^2) = {\Gamma (p+1) E^{p-3/2} \over (2\pi)^{3/2} \Gamma (a+b)}
\curlyH (a,b,p-\frac{1}{2},1; \frac{L^2}{2E}),\eqno\new$$
where
$$\curlyH(a,b,c,d;x) = {1\over 2\pi i} \int_{\cal C} {\Gamma (a+s)
\Gamma (b-s)\over \Gamma (c+s) \Gamma (d-s)} x^{-s} ds. \eqno\new$$
Here, ${\cal C}$ is a contour in the complex $s$ plane such that all
the poles $-a-n$ are on the left and all the poles $b+n$ are on the
right (where $n$ is an integer).  Dejonghe (1986) shows how to
evaluate this integral. When $x \le 1$, the contour is completed to
the left by adding a large semi-circle at infinity. Only the poles
at $s = -a-n$ are enclosed, giving
$$\curlyH(a,b,c,d;x)\!=\!{\Gamma (a\!+\!b) x^a \over \Gamma (c\!-\!a) 
\Gamma (a\!+\!d)}{}_2F_1(a\!+\!b, 1\!+\!a\!-\!c, a\!+\!d; x).$$
When $x > 1$, the contour is completed to the right and only the poles
at $s = b+n$ contribute to give
$$\curlyH(a,b,c,d;x)\!=\!{\Gamma (a\!+\!b) x^{-b} \over \Gamma (d\!-\!b) 
\Gamma (b\!+\!c)}{}_2F_1(a\!+\!b, 1\!+\!b\!-\!d, b\!+\!c; {\displaystyle 1 \over
\displaystyle x}).$$

Second, let us extend this result to the models with rising rotation
curves ($\alpha<0$). In this case, the Fricke components are derived
by Evans (1994) as
$$\rho' = {1\over (-\psi)^p} r^{-\beta},\eqno\new$$
$$f' = {\Gamma (p-\beta/2 + 3/2)\over \Gamma( 1- \beta/2)\Gamma(p)}
{L^{-\beta}\over (-E)^{p-\beta/2 + 3/2}  (2\pi)^{3/2}
2^{-\beta/2}}.\eqno\new$$
Again, we seek the DF $F$ corresponding to the density
$$\rho = {1\over (-\psi)^p} {r^{2a} \over (1+r^2)^{a+b}}.\eqno\new$$
This time the result is
$$\eqalign{F(E,L^2) =& {1\over (2\pi)^{3/2} \Gamma(p)\Gamma (a+b)} \cr
&\times {1 \over (-E)^{p+3/2}}
\curlyG (a,b,p+\frac{3}{2},1; \frac{L^2}{2E}),}\eqno\new$$
where
$${\curlyG} (a,b,c,d;x) = {1\over 2\pi i} \int_{\cal C}
{\Gamma (a\!+\!s) \Gamma (b\!-\!s) \Gamma (c\!-\!s) 
\over \Gamma (d\!-\!s)} (-x)^{-s} ds.
\eqno\new$$
Here, ${\cal C}$ is a contour in the complex $s$ plane such that all
the poles $-a-n$ are on the left and all the poles $b+n$ and $c+n$
are on the right (where $n$ is an integer). When $|x| \le 1$, we obtain
$${\cal G}(a,b,c,d;x) \!=\!{\Gamma (a\!+\!b)\Gamma (a\!+\!c)(-x)^a \over 
\Gamma (a\!+\!d)}{}_2F_1(a\!+\!b,\!a\!+\!c, a\!+\!d; x).$$
When $|x| > 1$, the contour is completed to the right and there are
two infinite sequences of poles at $b+n$ and $c+n$. We therefore
obtain a sum of two hypergeometric functions, namely
$$\eqalign{{\cal G}(a,b,c,d;x) =& {\Gamma (c\!-\!b)\Gamma
(a\!+\!b) \over (-x)^b
\Gamma (d\!-\!b)}{}_2F_1(a\!+\!b,1\!+\!\!b\!-\!d, 1\!+\!b\!-\!c; 
\frac{1}{x}) \cr
+& {\Gamma (b\!-\!c)\Gamma (a\!+\!c) \over (-x)^c
\Gamma (d\!-\!c)}{}_2F_1(a\!+\!c,1\!+\!\!c\!-\!d, 1\!+\!c\!-\!b; 
\frac{1}{x}).\cr}$$

The third and final case is that corresponding to a flat rotation
curve ($\alpha =0$). The elementary Fricke components become
$$\rho' = r^{-\beta/2}\exp (p\psi/v_0^2),\eqno\new$$
and
$$f' = {p^{3/2-\beta/2} \over (2\pi)^{3/2} 2^{-\beta/2}\Gamma (1-
\beta/2) v_0^{3-\beta/2}} L^{-\beta/2} \exp( pE/v_0^2 ).
\eqno\new$$
This means that the density
$$\rho = {r^{2a} \over (1\!+\!r^2)^{a\!+\!b}} \exp (p\psi/v_0^2),
\eqno\new$$
corresponds to the DF 
$$F ={p^{3/2+a}\over (2\pi)^{3/2} 2^a \Gamma (a\!+\!1) v_0^{2a+3}}
\exp(pE/v_0^2)\Phi(a\!+\!b,a\!+\!1, -\frac{pL^2}{2v_0^2}),\eqno\new$$
where $\Phi$ is the degenerate hypergeometric function.

\section{Numerical Computation of Hypergeometric Functions}

The hypergeometric function is defined within the unit circle
$z<1$ by the hypergeometric series
$$
\eqnam{\hypseries}
\eqalign{
&_2F_1(a,b;c;z) = 1 + {ab\over c\, 1!} z\!+\!{a (a\!+\!1)\, b
(b\!+\!1)\over c (c\!+\!1) \,2!} z^2 + \ldots  \cr
&+ {a (a\!+\!1)\ldots (a\!+\!n\!-\!1)\, b (b\!+\!1)\ldots 
(b\!+\!n\!-\!1)\over c(c\!+\!1)\ldots (c\!+\!n\!-\!1) \, n!} 
z^{n}\!+\!\ldots}\eqno\new$$
As the coefficients of the series approach 1 as $n\rightarrow\infty$,
the series is bounded by a majorizing geometric series, and converges
for $|z|<1$.  Outside of the unit circle, an analytic continuation of
the hypergeometric series is given by the hypergeometric differential
equation
$$
\eqnam{\hypode}
_2F_1^{\prime\prime} = { ab \,_2F_1 - 
[c\!-\!(a\!+\!b\!+\!1)z]\,_2F_1^\prime \over z  (1\!-\!z)} \eqno\new
$$
A second, linearly independent solution to the hypergeometric
differential equation is $z^{1-c}\, _2F_1(a-c+1,b-c+1;2-c;z)$
(provided $c$ is not an integer).

Numerical evaluation of the hypergeometric function can be awkward.
In fact, none of the standard reference books on numerical methods
(e.g., Press et al. 1992) present completely general algorithms for
the computation of the hypergeometric function valid for all parameter
values. Luke (1977) has derived several useful Chebyshev and rational
polynomial approximations.  The coefficients of the hypergeometric
series (and its approximations) may reverse sign, so that the final
value can depend on the near-cancellation of very large terms,
rendering the normal fifteen digit floating point precision of a
computer insufficient.  The parasitic secondary solution of the
differential equation, by virtue of its steep $z^{1-c}$ growth,
renders the integration of the differential equation numerically
impracticable for $c \ll 0$.

Our algorithm divides the evaluation of the hypergeometric function
into two r\'egimes.  Provided $c \gta -3$, the hypergeometric
differential equation (\hypode) is stable. It is integrated via the
standard ordinary differential equation integration methods provided
by Press et al (1992). The initial condition for the integration
cannot be chosen as ${}_2F_1(a,b,c;z) = 1$ when $z=0$, as the
differential equation (\hypode) is then singular. Accordingly, we
start from $z = \epsilon$ and calculate the value of
${}_2F_1(a,b,c;z)$ at this point by direct summation of the
hypergeometric series. This is the method of choice for evaluation of
the DF when $\alpha >0$ or when $\alpha <0$ and $|L^2/(2E)| \le 1$.

If $c<-3$ (as occurs in the expressions for the DF when $\alpha<0$ and
$|L^2/(2E)| >1$) and $z > -0.5$, we compute the power series using
standard double precision arithmetic, but switch to arbitrary
precision arithmetic whenever the terms of the series exceed $10^6$,
ensuring that the final value will be accurate to nine decimal places.
Arbitrary precision arithmetic is necessary only for
$\alpha\rightarrow0$, or $c\ll 0$. If $z < -0.5$ then we
use the transformation (see eq. [9.131.1] of Gradshteyn \& Ryzhik
1978)
$$
_2F_1(a,b;c;z)=(1-z)^{-a}\,_2F_1\left(a,c-b;c;z/(z-1)\right)\eqno\new
$$
to cast $z$ into $z/(z-1)$, ensuring that the power series converges
swiftly for all $z$ of interest.

\bye